\documentclass[aps,twocolumn,showpacs,amsmath,amssymb,pra,superscriptaddress,floatfix,longbibliography]{revtex4-1}

\usepackage{bm}
\usepackage{physics}
\usepackage{braket}
\usepackage{amsmath}
\usepackage{amssymb}
\usepackage{mathtools}
\usepackage{graphicx}
\usepackage{dcolumn}
\usepackage{bm}
\usepackage{multirow}
\usepackage{leftidx}
\usepackage{color}
\usepackage{float}
\usepackage{capt-of}
\usepackage{mwe}
\usepackage{amsfonts}
\usepackage[capitalise]{cleveref}
\usepackage[shortlabels]{enumitem}
\usepackage[german,english]{babel}
\usepackage{gensymb}
\usepackage[thinc]{esdiff}
\usepackage{url}
\graphicspath{ {./Images/Free}, {./Images/bound} }
\setlist[enumerate,1]{label={(\roman*)}}

\begin{document}


\title{Simulations for x-ray imaging of wave-packet dynamics}


\author{Akilesh Venkatesh}
\affiliation{Department of Physics and Astronomy, Purdue University, West Lafayette, Indiana 47907, USA}

\author{F. Robicheaux}
\affiliation{Department of Physics and Astronomy, Purdue University, West Lafayette, Indiana 47907, USA}


\date{\today}

\begin{abstract}
Previous work on imaging wave packet dynamics with x-ray scattering revealed that the scattering patterns deviate substantially from the notion of instantaneous momentum density of the wave packet. Here we show that scattering patterns can provide clear insights into the electron wave packet dynamics if the final state of the scattered electron and the scattered photon momentum are determined simultaneously. The scattering probability is shown to be proportional to the modulus square of the Fourier transform of the instantaneous electronic spatial wave function weighted by the final state of the electron. Several cases for the choice of final state of the electron are explored. First, the case where the final state can be measured up to a given principal quantum number $n$ and orbital angular momentum $l$ are presented. Next, the case where the final states can only be determined up to a given energy is discussed. Finally, the case of an initial wave packet consisting of a large amount of a known stationary state and a small amount of an unknown stationary state is examined. The scattering profile is used to determine the properties of the unknown state in the wave packet.   
\end{abstract}

\pacs{}

\maketitle


\section{Introduction}
Since the discovery of Bragg diffraction~\cite{braggdiffraction1913} and Compton scattering~\cite{Compton} over a century ago, x rays have been used as a probe at the subatomic scale ~\cite{Xray_Review_RIXS, Xray_diffraction_microscopy, ultrafast_xray_spectroscopy, coherent_Xrays_nugent2010}. X rays generally exhibit a low scattering cross section with matter. Therefore, as x rays travel through a sample, the likelihood of re-scattering after the first scattering event is minimal, making them a useful probing tool. For decades, the scattering of x rays from an electron in a stationary state has been used to map the momentum density of the electron, also known as the Compton profile~\cite{Comptonprofile, Comptonprofile2, CP_1986_theoryexp_check, CP_experimental_EMD}. Such measurements on a system in a stationary state are independent of time. However, measurements on a system described by a wave packet, which is a superposition of stationary states, will generally depend on time. To study the time dynamics of electronic wave packets, one requires pulses that have a pulse duration comparable to the oscillatory timescale of the wave packet. In the last two decades, the advent of x-ray free-electron lasers (XFEL)~\cite{XFEL1, LCLS_5yrs, SACLA_1, SACLA_2, EuropeanXFEL1, EuropeanXFEL2, EuropeanXFEL3} that generate pulses in the femtosecond and now attosecond timescales has made this possible. Pump-probe experiments at the attosecond timescale outside the x-ray regime have been previously shown to be quite effective in studying both atomic and molecular wave packets~\cite{pumpprobe_atto_valenceelectron,pumpprobe_atto_molecule_2, kim2012_PRL_IRprobe}. The measurement of time-dependent electron density in molecules can be used to construct molecular movies that offer insight into molecular processes such as bond formation and breaking~\cite{itatani2004tomographic, bondbreaking_stuff_dixit, xrayultrafast_moleculardynamics_2018, Mukamel_Xraydiffraction_2018, Dixit_benzene_molecule}.

While x-ray scattering from an electronic stationary state can be used to access the momentum density of the electron, the results for x-ray scattering from a wave packet was shown to have non-trivial dependence on the instantaneous charge density of the electron~\cite{Dixit_mainpnas}. In 2012, Dixit et al.~\cite{Dixit_mainpnas} showed that when x rays scatter from an electron in a wave packet state, the incident x-ray field can inelastically scatter, causing transitions from the wave packet state to several final states which depend on the bandwidth of the x-ray pulse. This leads to scattering patterns that deviate substantially from the Fourier transform of the instantaneous charge density of the wave packet. Some of the works that have followed Ref.~\cite{Dixit_mainpnas} have offered alternative techniques to extract information about the instantaneous charge density of the wave packet~\cite{Dixit_proposed_PCI_PRL, Santra_offresonant_term, Santra_finalecontinuum}. For instance, in Ref.~\cite{Dixit_proposed_PCI_PRL}, Dixit et al. describe a phase contrast imaging technique by examining the interference between the incident and scattered field to obtain the Laplacian of the projected instantaneous charge density. However, this approach requires the placement of detectors in the near field regime which may be experimentally challenging. Recently, Grosser et al.~\cite{Santra_finalecontinuum} showed that using inelastic Compton scattering, one can achieve x-ray imaging of electron wave packet dynamics provided the scattered electron ends up in a continuum state.

It should be noted that theoretical descriptions of time-resolved x-ray scattering have existed prior to Ref.~\cite{Dixit_mainpnas} and the first fully-quantized description of this problem can be attributed to Henriksen and Møller~\cite{henriksen2008theory}. A detailed overview of the history of the theoretical descriptions can be found in Ref.~\cite{Sim_2019}. In this paper, we derive the double differential scattering probability for x rays to scatter from an electron in a non-stationary state into a specified final state (or states) resulting in an expression which is related to the ones in Ref.~\cite{henriksen2008theory, Dixit_mainpnas, Sim_2017, Sim_2019}. We show that if the final state of the electron after scattering can be detected, it is possible to obtain meaningful information about the dynamics of the electronic wave packet. The scattering profile is shown to reveal the modulus square of the Fourier transform of the instantaneous transition charge density of the electron.

The paper is organized in the following manner: In Sec.~\ref{Methods}, the double differential scattering probability for x-ray scattering from an electron wave packet is derived and the expression is tailored for the special case of an electron wave packet made of two eigenstates. In Sec.~\ref{Sec_applications}, the imaging technique is illustrated using several examples. In Sec.~\ref{conclusion_summary}, the conclusions and a summary of the paper are presented.

Unless otherwise stated, atomic units will be used throughout this work.

\section{Methods and modelling} \label{Methods}
\subsection{Deriving the double differential scattering probability}
The approach we use to model x-ray scattering from electrons involves treating the incoming field classically and quantizing the outgoing field~\cite{KB}. Since the problem is non-relativistic for the parameter regime explored, a time-dependent Schr\"odinger equation approach is adequate.  One-photon scattering processes in the parameter regime studied in this paper were shown in Ref.~\cite{NLC_interference} to be adequately described by the first order perturbative treatment of the incoming field. For a complete description of this approach and its validity, see Ref.~\cite{NLC_interference}. Using this approach, the scattering probability amplitude for the process $\psi_{\boldsymbol{k},\boldsymbol{\epsilon}} ^ {(1)}$ can be obtained from the following equation: 

\begin{equation} \label{Inhomogeneous_first}
\begin{split} 
  i \frac{\partial \psi_{{\boldsymbol{k},\boldsymbol{\epsilon}}} ^ {(1)} }{\partial t} - \hat{H}_{a} \psi_{\boldsymbol{k},\boldsymbol{\epsilon}} ^{ (1) } =  &\sqrt{\frac{2\pi}{ V\omega_{k}} } e^{-i\boldsymbol{k\cdot r} }  e^{i\omega_{k} t }    \\    
  & \times \boldsymbol{\epsilon}^* \cdot ( \boldsymbol{\hat{P}} ~ \psi_1^{(0)} + \boldsymbol{A_C} ~ \psi_0^{(0)} )~W(t) \\
  & + (\boldsymbol{A_C} \cdot \boldsymbol{\hat{P}}) \psi_0^{(1)}.
\end{split}
\end{equation}
The quantity $\psi_i ^ {(j)}$ refers to the scattering probability amplitude for a process that is of order $i$ in the incoming classical field and order $j$ in the outgoing quantized field. Note that the quantity $\psi_{{\boldsymbol{k},\boldsymbol{\epsilon}}}^{(1)}$ in Eq.~(\ref{Inhomogeneous_first}) corresponds to $\psi_1 ^ {(1)}$ in this notation~\cite{NLC_interference}. In Eq.~(\ref{Inhomogeneous_first}), $\hat{H}_{a}$ is the Hamiltonian for an electron in the absence of the incident x-ray field, $V$ is the quantization volume, and $\boldsymbol{k}$ and $\omega_k$ are the scattered photon momentum and angular frequency respectively. $\epsilon$ denotes the scattered photon polarization. Here, $\boldsymbol{k} \cdot \boldsymbol{\epsilon} = 0$ and $\omega_{k}$ = \boldsymbol{$|k|$}$c$ with $c$ being the speed of light in vacuum ($\sim$137.036 a.u.). $\boldsymbol{\hat{P}}$ is the canonical momentum operator for the electron, $\boldsymbol{A_C}$ is the vector potential for the classical incoming field, $\boldsymbol{r}$ refers to the position vector associated with the electron and $t$ refers to time. $W(t)$ is a windowing function which turns the source terms on only for the duration of the incident x-ray pulse. The final results are independent of the choice of the windowing function provided it is sufficiently smooth.~\cite{NLCPRA_1, NLC_interference}

For the parameter regime explored, the only source term in Eq.~(\ref{Inhomogeneous_first}) that contributes to the scattering probability is the Compton scattering term $\boldsymbol{A_C} \psi_0^{(0)}$~\cite{NLC_interference}. This term is sometimes referred to as the off-resonant contribution in x-ray scattering~\cite{Ac_approx_PRL_santrareference, Santra_offresonant_term} and this step is analogous to neglecting the dispersive correction term in Ref.~\cite{Dixit_mainpnas}.  The time dependence of $\psi_0^{(0)}$ is dictated by the time-dependent Schr\"odinger equation with the field-free Hamiltonian $\hat{H}_{a}$. The vector potential of the incoming pulse $\boldsymbol{A}_C$ is chosen to be the following:
\begin{equation}\label{classicalvectorpotential}
\begin{split} 
    \boldsymbol{A}_C = &\frac{E}{\omega_{in}} \cos\bigg[( \omega_{in} t - {\boldsymbol{k_{in}} \cdot \boldsymbol{r} } ) \bigg]  \\
    &\times \exp\Bigg[\frac{(-(2 \ln{2} ) (t - \frac{{\boldsymbol{\hat{k}_{in}} \cdot \boldsymbol{r} }}{c})^2)}{t^2_{wid}} \Bigg]  \boldsymbol{\epsilon_{in}} ,
\end{split}
\end{equation}
where $E$, $\omega_{in}$, $\boldsymbol{k_{in}}$, $t_{wid}$, and $\boldsymbol{\epsilon_{in}}$ refer to the incoming electric field amplitude, angular frequency, momentum, the full width at half maximum (FWHM) of the pulse intensity, and polarization of the incoming field respectively.
The interaction between the electron and the incident x-ray field is modelled by considering the full space and time dependence of the vector potential because in the studied parameter regime the dipole approximation has limitations~\cite{Moe_Forre_ionization}. Using Eq.~(\ref{classicalvectorpotential}) and applying the rotating-wave approximation to Eq.~(\ref{Inhomogeneous_first}) yields
\begin{equation} \label{Inhomogeneous_first_rotatingwave}
\begin{split} 
  i \frac{\partial \psi_{\boldsymbol{k},\boldsymbol{\epsilon}} ^ {(1)} }{\partial t} - \hat{H}_{a} \psi_{\boldsymbol{k},\boldsymbol{\epsilon}} ^ {(1)} =  &\sqrt{\frac{2\pi}{ V\omega_{k}} } \frac{1}{2} \frac{E}{\omega_{in}} \boldsymbol{\epsilon}^* \cdot \boldsymbol{\epsilon}_{in} \\ 
  & \times \exp\Bigg[\frac{(-(2 \ln{2}) (t - \frac{{\boldsymbol{\hat{k}_{in}} \cdot \boldsymbol{r} }}{c})^2)}{t^2_{wid}} \Bigg] \\
  & \times e^{i(\boldsymbol{k_{in}} - \boldsymbol{k} )\cdot \boldsymbol{r} }  e^{i(\omega_{k} - \omega_{in}) t } ~  \psi_0^{(0)}.
\end{split}
\end{equation}

The scattering probability amplitude in the bra-ket notation can be expanded in an eigenbasis of electronic bound states and continuum states of the field-free Hamiltonian $\hat{H}_{a}$:
\begin{equation} \label{Expand_psi1_basis}
\ket{\psi_{\boldsymbol{k},\boldsymbol{\epsilon}} ^ {(1)} (t)} = \sum\limits_n C_{n,\boldsymbol{k} \boldsymbol{\epsilon} }(t) e^{-i {E_n} t} \ket{\psi_n }.
\end{equation}
Here \{$n$\} includes the set of all bound and continuum states and $E_n$ denotes the corresponding eigenenergies. 

Substituting Eq.~(\ref{Expand_psi1_basis}) in Eq.~(\ref{Inhomogeneous_first_rotatingwave}), one obtains the following expression for $C_{n, \boldsymbol{k} \boldsymbol{\epsilon}}$ in the bra-ket notation after integrating over $t \in (-\infty, \infty )$,
\begin{equation} \label{Cn_step1}
\begin{split} 
   \lim_{t\to\infty} C_{n, \boldsymbol{k} \boldsymbol{\epsilon}} (t)  =  & -i \sqrt{\frac{2\pi}{ V\omega_{k}} } \frac{1}{2} \frac{E}{\omega_{in}} \boldsymbol{\epsilon}^* \cdot \boldsymbol{\epsilon}_{in} 
   \int^{\infty}_{-\infty} dt \bra{\psi_n} e^{i {E_n} t} \\
   & \times \exp\Bigg[\frac{(-(2 \ln{2}) (t - \frac{{\boldsymbol{\hat{k}_{in}} \cdot \boldsymbol{r} }}{c})^2)}{t^2_{wid}} \Bigg] \\
  & \times e^{i(\boldsymbol{k_{in}} - \boldsymbol{k} )\cdot \boldsymbol{r} }  e^{i(\omega_{k} - \omega_{in}) t } ~  \ket{\psi_0^{(0)}}.
\end{split}
\end{equation}

In this paper, the initial state of the electron ($\ket{\psi_0^{(0)}}$) is described  by a wave packet. The wave packet can be expanded in the same basis as Eq.~(\ref{Expand_psi1_basis}), 
\begin{equation} \label{Expand_wpkt_basis}
\ket{\psi_0 ^ {(0)} (t)} = \sum\limits_{n''} a_{n''} ~e^{-i {E_{n''}} t} \ket{\psi_{n''} }.
\end{equation}
Here, $a_{n''}$ is the probability amplitude associated with state $\ket{\psi_{n''}}$ at $t = 0$. The envelope function of the incoming classical field [Eq.~(\ref{classicalvectorpotential})] can be approximated as a pure Gaussian since  ${\boldsymbol{\hat{k}_{in}} \cdot \boldsymbol{r} } / c << t_{wid}$. Then, Eq.~(\ref{Cn_step1}) after integration over time is
\begin{equation} \label{Cn_step2}
\begin{split} 
   \lim_{t\to\infty} C_{n, \boldsymbol{k} \boldsymbol{\epsilon}} (t)  =  & -i \sqrt{\frac{2\pi}{ V\omega_{k}} } \frac{1}{2} \frac{E}{\omega_{in}} t_{wid} \sqrt{\frac{\pi}{2 \ln{2} }} \boldsymbol{\epsilon}^* \cdot \boldsymbol{\epsilon}_{in} \\ 
   & \times \sum_{n''} a_{n''}
   e^{-\big( E_{n} - E_{n''} + \omega_{k} - \omega_{in} \big)^2 \frac{t^2_{wid}}{8 \ln{2}  } } \\
  & \times \bra{\psi_n} ~
  e^{i(\boldsymbol{k_{in}} - \boldsymbol{k} )\cdot \boldsymbol{r} }   ~  \ket{\psi_{n''}}.
\end{split}
\end{equation}
Given that the scattered electron is in the state $\ket{\psi_f}$, the double differential scattering probability is given by the modulus square of the corresponding scattering probability amplitude $C_{f,\boldsymbol{k} \boldsymbol{\epsilon}}$.

\begin{equation} \label{prob_step1}
\begin{split} 
    \pdv{P_f( \boldsymbol{Q} )}{\Omega~}{\omega_k} =  &  \frac{2\pi}{ V\omega_{k}} \frac{1}{4} \frac{E^2}{\omega^2_{in}} t^2_{wid} ~ \frac{\pi}{2 \ln{2} } |\boldsymbol{\epsilon}^* \cdot \boldsymbol{\epsilon}_{in}|^2 \\ 
   & \times \sum\limits_{ n', n'' } a^*_{n'} a_{n''} \bra{\psi_{n'}} e^{-i \boldsymbol{Q} \cdot \boldsymbol{r} }  \ket{\psi_{f}} \\
  & \times e^{-\big[(\epsilon_f - E_{n'})^2 + (\epsilon_f - E_{n''})^2 \big]  \frac{t^2_{wid}}{8 \ln{2}  }  } \\
  & \times \bra{\psi_{f}} e^{i \boldsymbol{Q} \cdot \boldsymbol{r} }  \ket{\psi_{n''}} ,
\end{split}
\end{equation}
where $\epsilon_f = E_f + \omega_k - \omega_{in}$ and $\boldsymbol{Q} = \boldsymbol{k}_{in} - \boldsymbol{k}$. Here the quantity $E_{f}$ denotes the energy of the stationary state $\ket{\psi_{f}}$

It is useful to examine Eq.~(\ref{prob_step1}) for the case of a free electron starting in a momentum eigenstate, the quantity inside the summation on the right hand side of Eq.~(\ref{prob_step1}) is unity only when  $\ket{\psi_f}$ is the corresponding momentum eigenstate that is allowed by momentum and energy conservation after the momentum kick from the x-ray photon. In all other cases, the summation results in zero. Therefore, the free electron double differential scattering probability is given by

\begin{equation} \label{dds_prob_free_electron}
\begin{split} 
    \pdv{P_{f}( \boldsymbol{Q}, \boldsymbol{p_i})}{\Omega~}{\omega_k} = & \pdv{P_e}{\Omega~}{\omega_k} ~ \delta(\boldsymbol{p_f} - \boldsymbol{p_i} - \boldsymbol{Q}) ,
\end{split}    
\end{equation}
where $\boldsymbol{p_f}$ and $\boldsymbol{p_i}$ are the momentums that correspond to the final and initial electron momentum eigenstates respectively and $\pdv{P_e}{\Omega~}{\omega_k}$ is defined as

\begin{equation} \label{prob_free_electron}
    \pdv{P_e}{\Omega~}{\omega_k} =   \frac{2\pi}{ V\omega_{k}} \frac{1}{4} \frac{E^2}{\omega^2_{in}} t^2_{wid} ~ \frac{\pi}{2 \ln{2} } |\boldsymbol{\epsilon}^* \cdot \boldsymbol{\epsilon}_{in}|^2.
\end{equation}

To understand how Eq.~(\ref{prob_step1}) describes the wave packet dynamics, recall that the quantity $a_n$ was defined to be the probability amplitude at $t=0$ and the incident x-ray pulse attained its peak intensity at $t=0$. For convenience, the peak intensity of the x-ray pulse is now shifted to a later time $t_0$ (delay time) by carrying out $t \rightarrow t - t_0$. Then, the double differential scattering probability for a bound electron as a function of the delay time can be written as

\begin{equation}\label{diffprob_mainexpression}
\begin{split}
  \pdv{P_f( \boldsymbol{Q}, t_0 )}{\Omega~}{\omega_k} = & \pdv{P_e}{\Omega~}{\omega_k}  \sum\limits_{ n', n'' } a^*_{n'} a_{n''} ~e^{i(E_{n'} - E_{n''} ) t_0} \\
  & \times \bra{\psi_{n'}} e^{-i \boldsymbol{Q} \cdot \boldsymbol{r} }  \ket{\psi_{f}}  \\
   & \times e^{-\big[(\epsilon_f - E_{n'})^2 + (\epsilon_f - E_{n''})^2 \big]  \frac{t^2_{wid}}{8\ln{2}}  } \\
  & \times \bra{\psi_{f}} e^{i \boldsymbol{Q} \cdot \boldsymbol{r} }  \ket{\psi_{n''}}.
\end{split}  
\end{equation}  
Note that Eq.~(\ref{diffprob_mainexpression}) does not depend explicitly on the electric field amplitude or the polarization of the incoming or outgoing field. The dependence on the polarization and electric field is however implicitly contained in $\pdv{P_e}{\Omega}{\omega_k}$. Therefore, it is convenient to scale Eq.~(\ref{diffprob_mainexpression}) by $\pdv{P_e}{\Omega}{\omega_k}$ to obtain the scaled double differential probability. The results from the final expression in Eq.~(\ref{diffprob_mainexpression}) has been evaluated and bench marked with the results of the non-perturbative Schr\"odinger equation from Ref.~\cite{NLCPRA_1, NLC_interference}. They show excellent agreement for the parameter regime discussed in the manuscript thus validating the approximations involved. 



 While expressions similar to Eq.~(\ref{diffprob_mainexpression}) have been derived previously~\cite{Dixit_mainpnas, Sim_2017, Sim_2019}, we have included a detailed derivation in this work to provide clarity and to discuss the different stages in the derivation where approximations are used. Equation~(\ref{diffprob_mainexpression}) differs from the expression given by Dixit et al.~\cite{Dixit_mainpnas} in two aspects. First, instead of the mean energy of the wave packet, the individual energies of the constituent stationary states ($E_n$) appear. This difference is only introduced towards the end of the derivation in Ref.~\cite{Dixit_mainpnas} and is a valid approximation if the x ray pulse width is much shorter than the oscillation period of the wave packet. Second, instead of a summation over all possible final states for the scattered electron, the final state is selected to be $\ket{\psi_f}$. In Ref.~\cite{Dixit_mainpnas} and other previous works~\cite{Sim_2017, Sim_2019}, it is the summation over all the final scattered electron states that makes it difficult to access the information about the instantaneous charge density of the wave packet. The lack of summation in Eq.~(\ref{diffprob_mainexpression}) allows one to extract the Fourier transform of the weighted probability amplitude of the instantaneous wave packet. This is similar to the idea implicit in Grosser et al.~\cite{Santra_finalecontinuum} where the final state of the electron was assumed to be a plane wave. The approach described in this paper however is not restricted to the case of a continuum state for the scattered electron but rather on the principle that determining the final state of the electron simultaneously with the scattered photon momentum allows access to the momentum density of the electron wave packet.


\subsection{Two-state wave packet}
While Eq.~(\ref{diffprob_mainexpression}) is valid for an arbitrary electronic wave packet, for simplicity we have used wave packets consisting of two eigen states for the derivation and examples below. There is no fundamental requirement for two states. The basic features are unchanged as long as the time scale of the wave packet is longer than that of the incident x-ray pulse.

Let the wave packet consisting of two eigenstates be,
\begin{equation}\label{wpkt_defn}
  \ket{\psi(t_0)} = a_\alpha\ket{\psi_{\alpha}} + a_\beta e^{i\phi(t_0)}\ket{\psi_{\beta}}.
\end{equation}
The instantaneous phase $\phi(t_0)$ satisfies $\phi(t_0) = (E_{\alpha} - E_{\beta}) * t_0$, where $E_{\alpha}$ and $ E_{\beta}$ are the eigenenergies corresponding to the eigenstates $\ket{\psi_{\alpha}}$ and $\ket{\psi_{\beta}}$ respectively. For such a wave packet one can then simplify Eq.~(\ref{diffprob_mainexpression}) by imposing a condition on the energy of the scattered photons~($\omega_k$). The condition is that the energy difference between scattered and incident xray, $\omega_k - \omega_{in}$, corresponds to the average transition energy between the stationary states in the wave packet and the final state of the electron ($\psi_f$). Then using Eq.~(\ref{diffprob_mainexpression}) the scaled double differential scattering probability can be written as 
\begin{equation}\label{wt_FT_equation}
\begin{split}
  \bigg( \pdv{P_f( \boldsymbol{Q}, t_0 )}{\Omega~}{\omega_k} \bigg)_{sc} = & ~ e^\frac{-\Delta E^2 t_{wid}^2}{16\ln{2}}  \\ 
  & \times \bigg| \int \psi^*_f(\boldsymbol{r}) ~ \psi(\boldsymbol{r},t_0) e^{i \boldsymbol{Q} \cdot \boldsymbol{r} } d^{3}r  \bigg|^2 ,
\end{split}  
\end{equation}
where $\Delta E$ is the energy difference between the two stationary states that constitute the wave packet~[Eq.~(\ref{wpkt_defn})]. The term $\psi_f(\boldsymbol{r})$ is the probability amplitude in position space of the final state of the electron. The quantity $\psi(\boldsymbol{r}, t_0)$ refers to the instantaneous probability amplitude of the wave packet in the absence of the incident x-ray field. It is evident from Eq.~(\ref{wt_FT_equation}) that the scaled double differential scattering probability is proportional to the Fourier transform of the instantaneous wave function of the electron weighted by the electron's final state wave function. Given the constraint on the scattered photon frequency, note that Eq.~(\ref{wt_FT_equation}) is identical to Eq.~(\ref{diffprob_mainexpression}) for a wave packet made of two eigenstates. It is worth pointing out that in some communities~\cite{Mukamel_TCD_defn, Mukamel_TCD_2, Mukamel_Eresolved, Mukamel_SXRI_2020} what we call as the weighted probability amplitude of the instantaneous wave packet in  Eq.~(\ref{wt_FT_equation}), that is $\psi^*_f(\boldsymbol{r}) \psi(\boldsymbol{r},t_0)$, is referred to as the transition charge density~\cite{Mukamel_TCD_defn}.

The result in Eq.~(\ref{wt_FT_equation}) is suitable for the ideal case where the detector resolution is assumed to be much smaller than the bandwidth of the  x-ray pulse. 
Experimentally in cases with limited detector resolution, one might be interested to integrate Eq.~(\ref{wt_FT_equation}) over a range of scattered photon energies over which the detector is sensitive, to obtain the differential scattering probability. A discussion of the energy-integrated double differential scattering probability and its implications are presented in Sec.~\ref{section_DDS_DS}.
 

\subsection{Convergence} \label{Sec_convergence}
To evaluate the double differential scattering probability, the relevant matrix elements in Eq.~(\ref{diffprob_mainexpression}) were calculated numerically on a grid in spherical coordinates. The convergence is determined by calculating the change in the scaled double differential scattering probability. The only convergence parameters that give rise to a measurable difference in the final results for the scaled differential probability are the radial grid spacing and the radial grid size.

\begin{figure}
\resizebox{87mm}{!}{\includegraphics{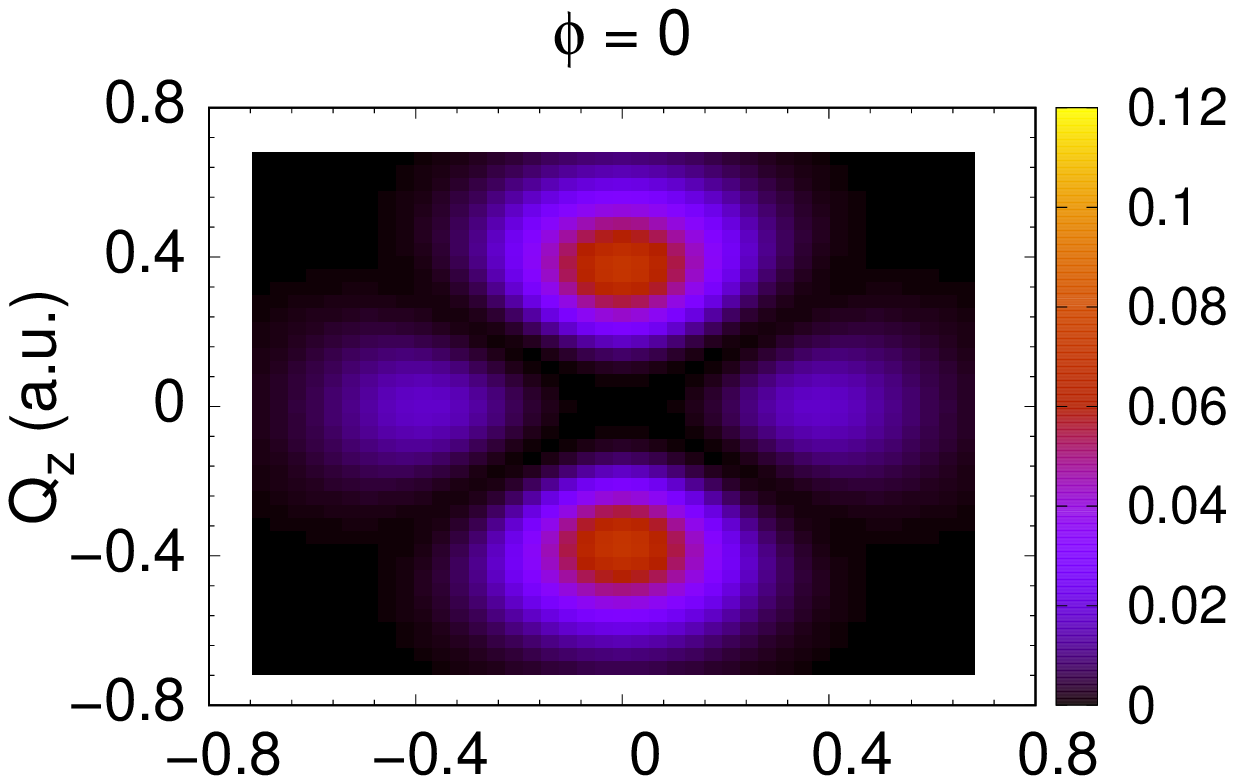}
\includegraphics{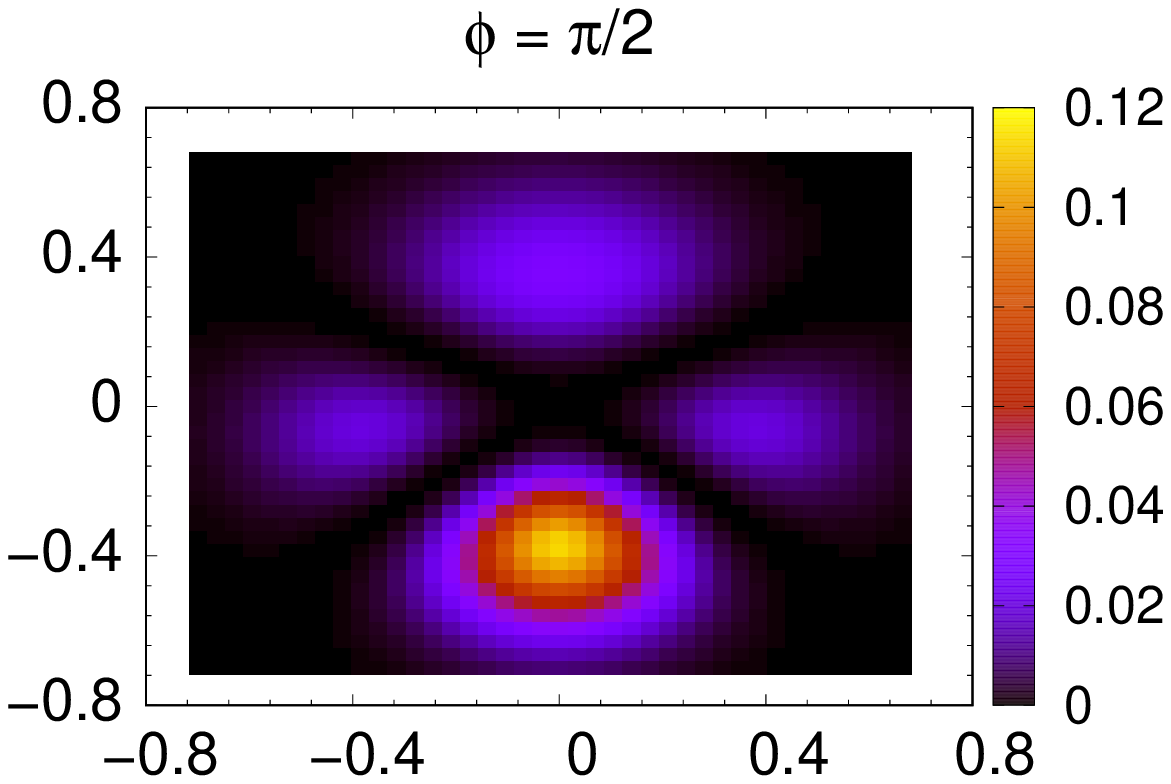}}
\resizebox{87mm}{!}{\includegraphics{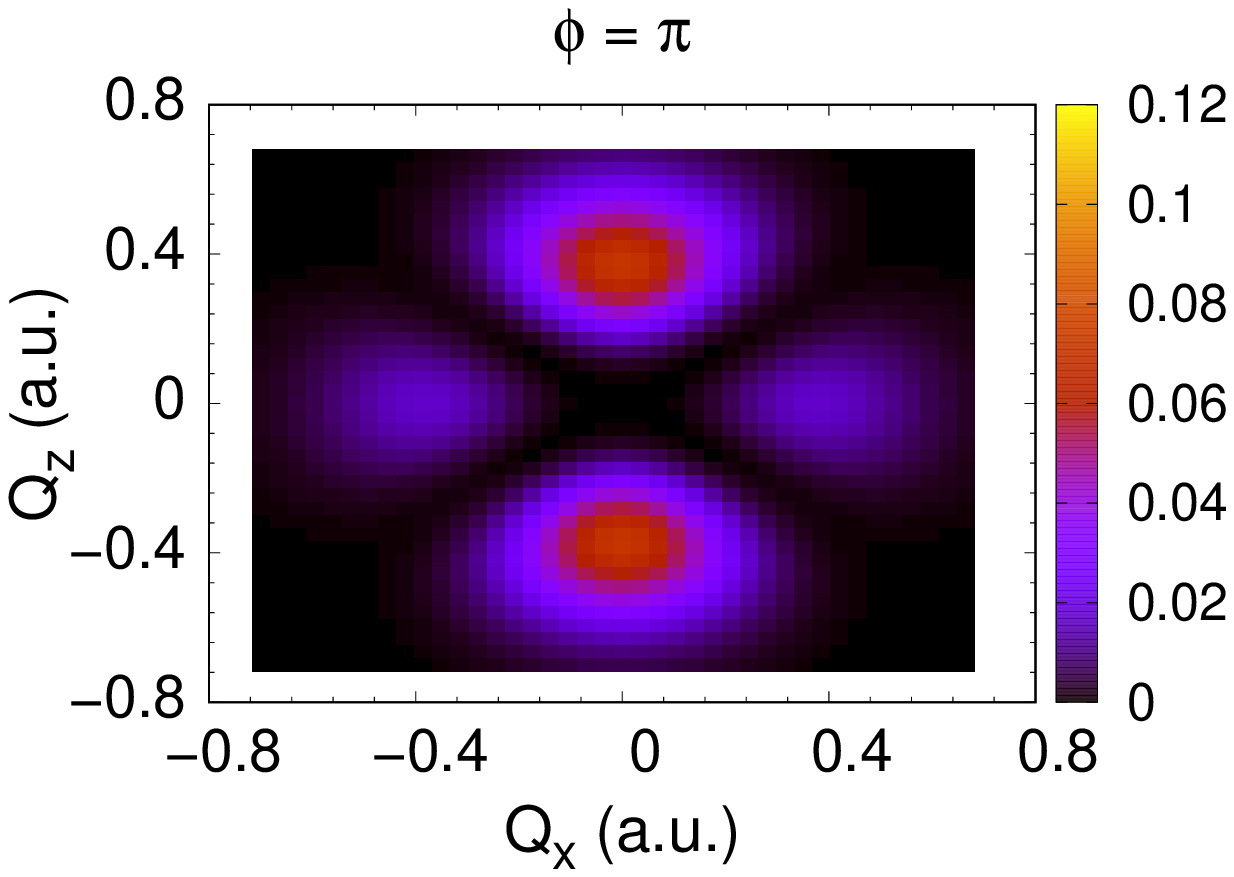}
\includegraphics{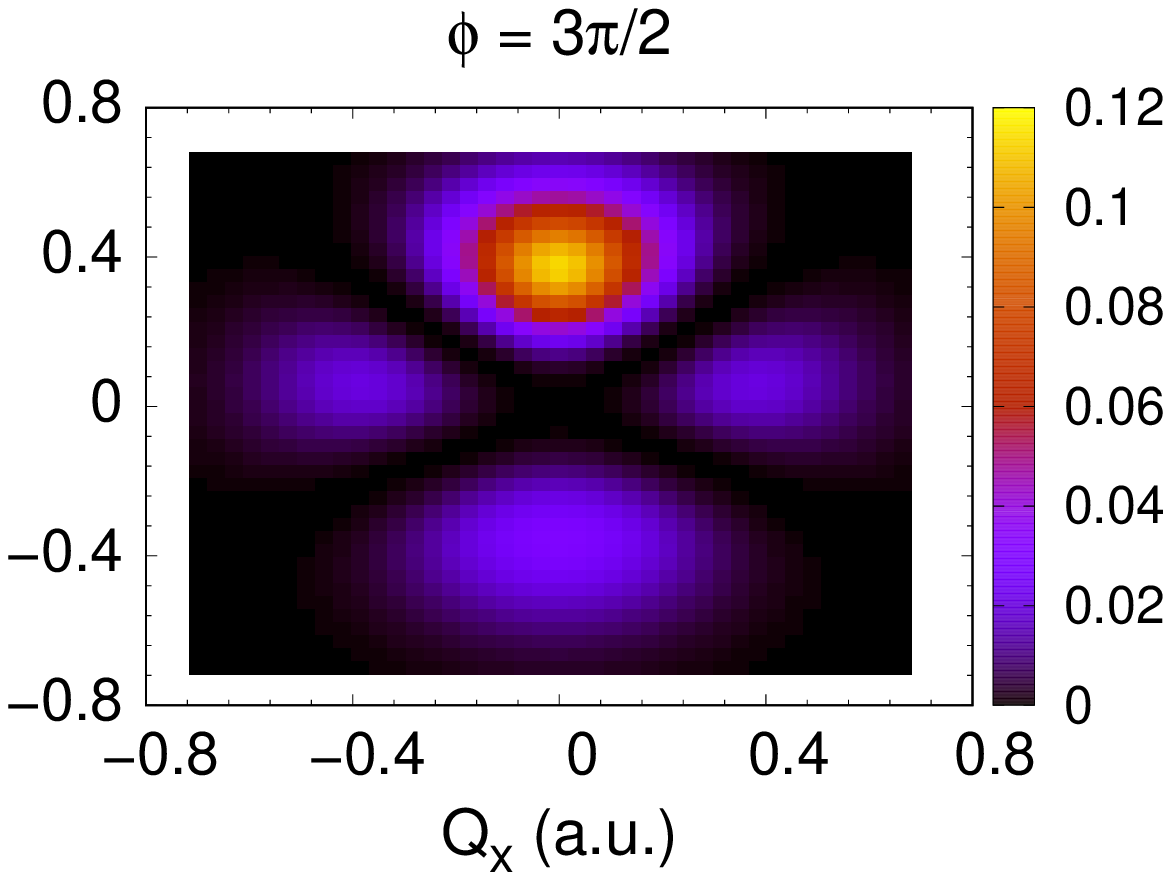}}
\caption{\label{Fig_Fix_mf}
The plots show the scaled double differential scattering probability vs momentum transferred to the electron in hydrogen calculated at different propagation times for the wave packet. The initial wave packet consists of equal probabilities of 3d and 4f, m=0 states. The different phase angles specified at the top of each tile correspond to different delay times~[Eq.~(\ref{wpkt_defn})] for the probe pulse. The final state of the scattered electron has been chosen to be 2$s$. The scaled double differential scattering probability is proportional to the modulus square of the Fourier transform of the instantaneous transition charge density~[Eq.~(\ref{wt_FT_equation})]. Here, $\omega_{in} = 147$ a.u.(4 keV), $t_{wid} = 41.34$ a.u. (1 fs).
}
\end{figure}

\begin{figure}
\resizebox{87mm}{!}{\includegraphics{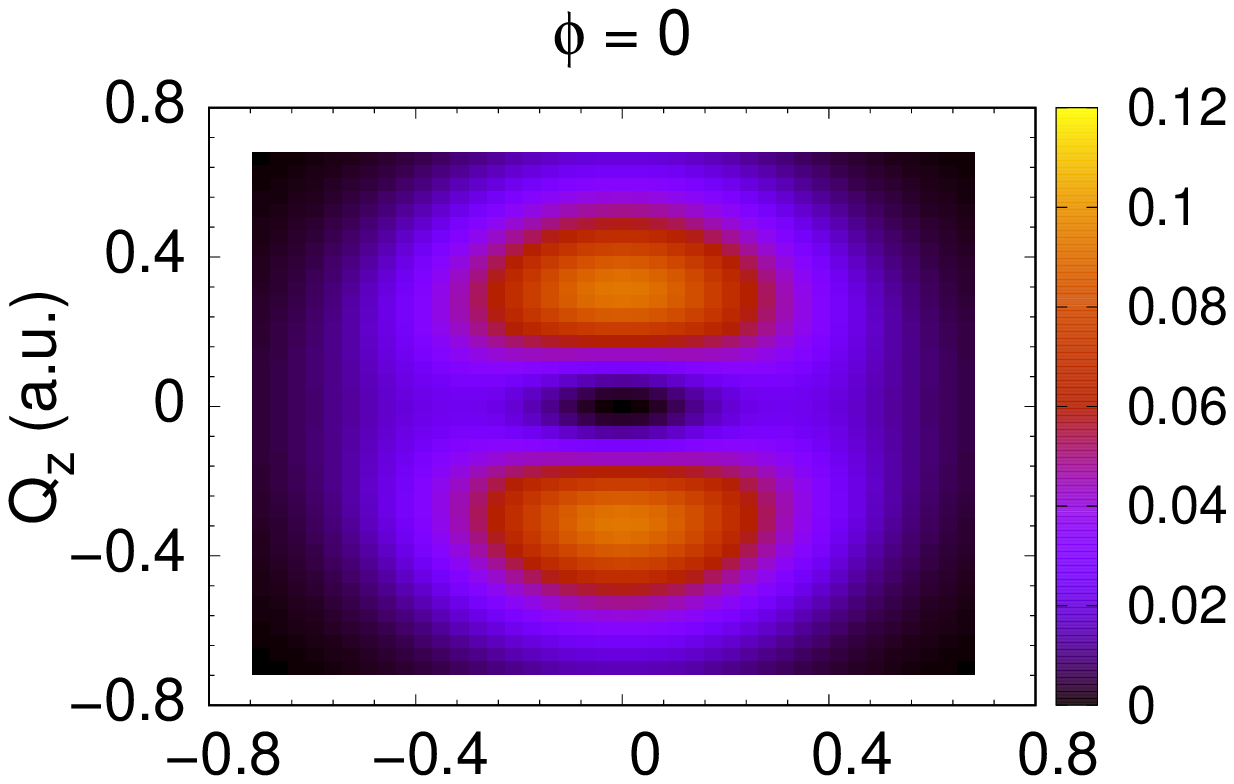}
\includegraphics{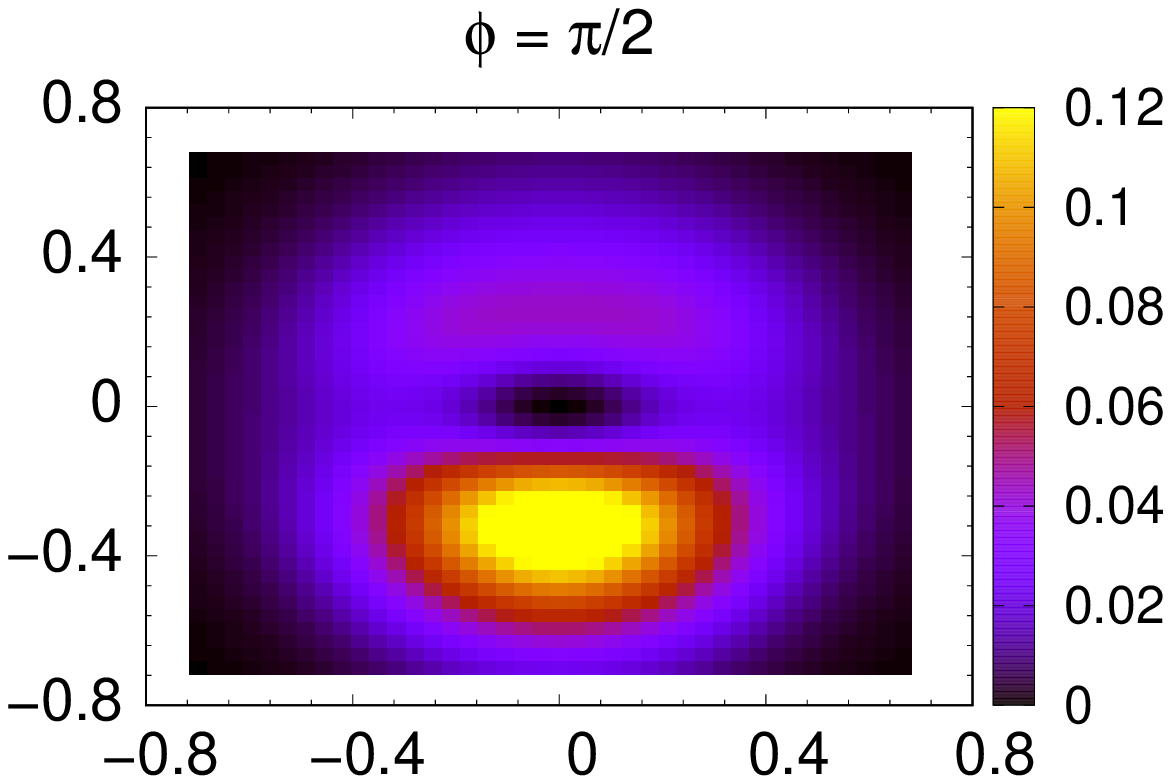}}
\resizebox{87mm}{!}{\includegraphics{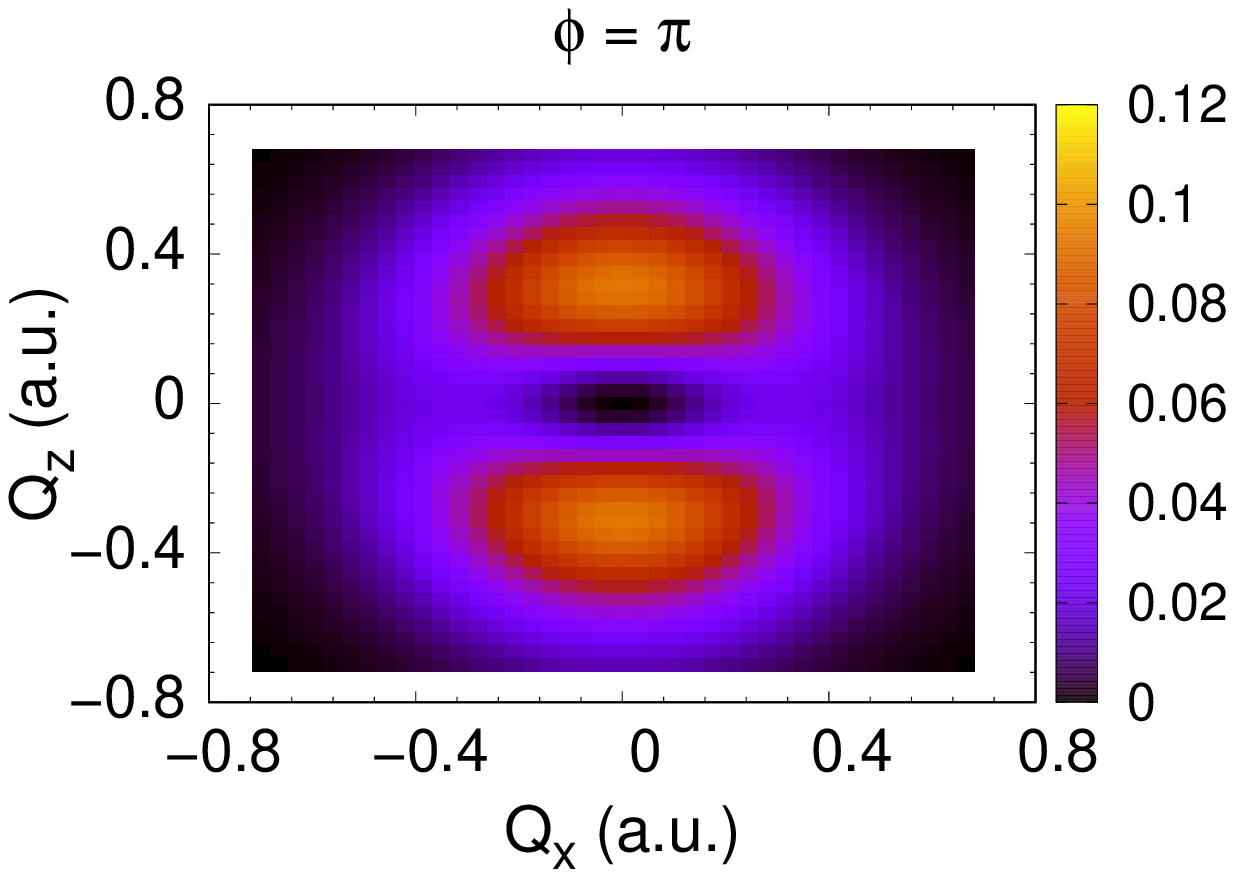}
\includegraphics{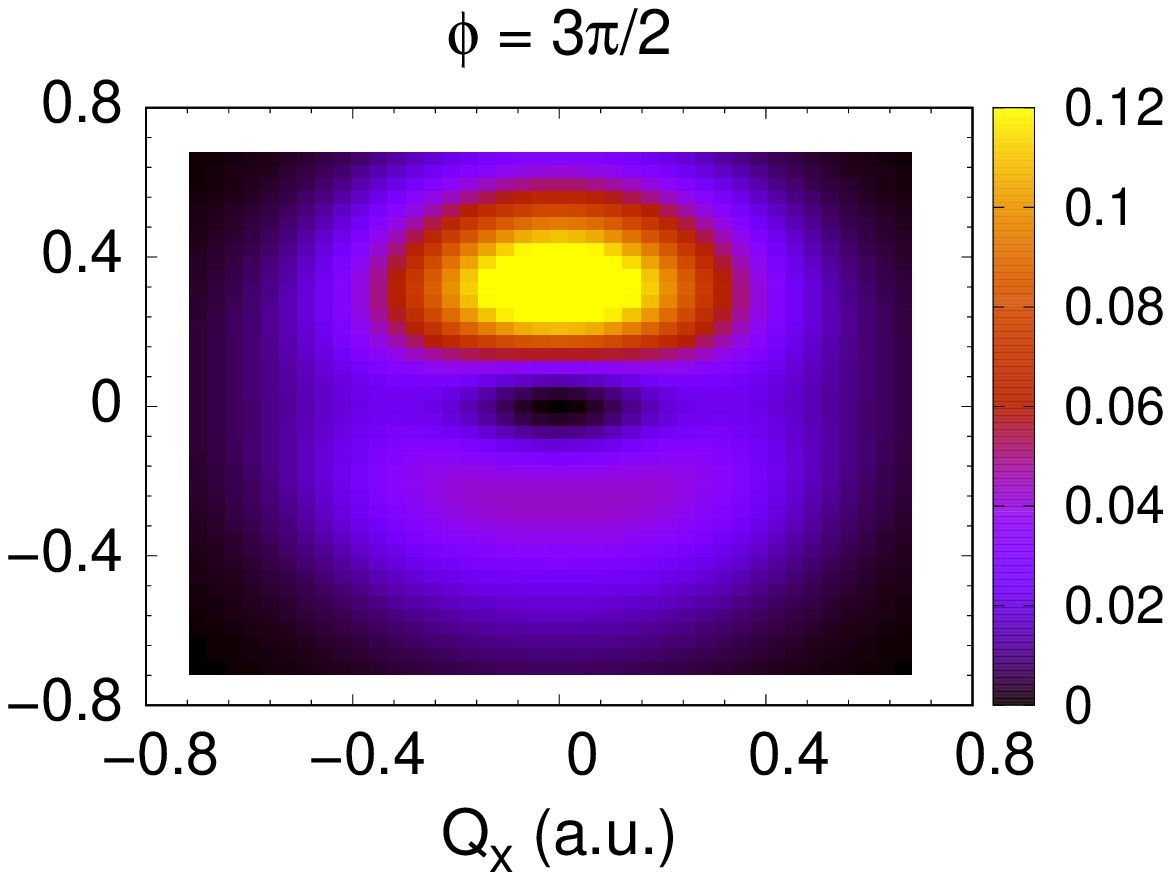}}
\caption{\label{Fig_sum_2s2p}
The results for the scaled double differential scattering probability vs momentum transferred to the electron calculated at different propagation times for the wave packet. The parameters are the same as that of Fig.~\ref{Fig_Fix_mf}, except here the results are summed over the electron final states 2$s$ and 2$p$ for all possible values of m. Even when the final state of the scattered electron can only be distinguished broadly based on the energy, dynamical phase information of the wave packet is still preserved.
}
\end{figure}

For the results in Fig.~\ref{Fig_Fix_mf}, a grid spacing of 0.1 and 0.05 a.u. leads to a difference of 0.81\% in the scaled double differential scattering probability. A grid size of 86 a.u. and 121 a.u. leads to difference of $10^{-6}$ \%. For the results in Fig.~\ref{Fig_sum_2s2p}, a grid spacing of 0.025 and 0.05 a.u.  gives a difference of 0.28 \% when the final state is 2s and 0.03 \% when the final state is 2p (summed over all m) respectively. The grid size parameters for Fig.~\ref{Fig_sum_2s2p} exhibit the same convergence behaviour as that of Fig.~\ref{Fig_Fix_mf} for both 2s and 2p final states. 

\section{Applications} \label{Sec_applications}

For the example calculations below, the wave packet is probed using an x-ray pulse with a mean photon energy of 147 a.u. (4 keV) and a pulse duration ($t_{wid}$) of 41.34 a.u.~(1 fs). The incoming field is chosen to be propagating in the $\hat{y}$ direction with its polarization in the $\hat{z}$ direction. It should be noted that Eq.~(\ref{diffprob_mainexpression}) does not make assumptions whether the system under study consists of atoms or molecules. However for a simple illustration of the method, we choose a wave packet that consist of two eigenstates of hydrogen.

\subsection{Fixing the final state up to a given $l$}
To illustrate this method, first we consider the case (Fig.~\ref{Fig_Fix_mf}) when the initial wave packet consist of equal weights of 3d and 4f m = 0 states of hydrogen. There exist several experimental techniques for preparing electronic wave packets (see Ref.~\cite{prepare_wpkt_bucksbaum1, prepare_wpkt_bucksbaum2} and references therein). In this paper, we begin our discussions by assuming that there exists a prepared electronic wave packet. 
This case corresponds to an instantaneous wave packet [Eq.~(\ref{wpkt_defn})] where $\ket{\psi_{\alpha}} = \ket{3, 2, 0}$, $\ket{\psi_{\beta}} = \ket{4, 3, 0}$ and $a_\alpha = a_\beta = \frac{1}{\sqrt{2}}$. Here $\ket{n,l,m}$ refers to a state described by the usual atomic quantum numbers $n$, $l$, and $m$ respectively.
 
In Fig.~\ref{Fig_Fix_mf}, the scaled double differential scattering probability is studied as the components of $\boldsymbol{Q}$ ($Q_x$ and $Q_z$ only) are varied independently. It should be noted that the component $Q_y$ is determined for a given $Q_x$ and $Q_z$ because of conservation of energy and momentum. The final state of the electron is chosen to be 2$s$. This is an ideal case when a final state with quantum numbers described by $n$ and $l$ can be precisely selected. This ideal case serves to provide a simple conceptual demonstration of the imaging technique. A similar example of an electronic wave packet in hydrogen has been discussed previously in Ref.~\cite{Dixit_mainpnas, Sim_2017} however in those previous works, the scattering pattern is the result of a summation over all possible final states and not for the case of a given final state. In this work, a numerical approach is used and convergent results have been obtained (see Sec.~\ref{Sec_convergence}) for scattering probabilities. Note that it has been shown by Ref.~\cite{Sim_2017} that for the case of hydrogenic wave packets it is possible to obtain analytic solutions if one employs parabolic coordinates.

A qualitative understanding of the scattering profile~(Fig.~\ref{Fig_Fix_mf}) can be obtained by looking at the different terms that contribute to the double differential scattering probability.
\begin{equation}\label{terms_wpkt_ddsp}
\begin{split}
\bigg( \pdv{P_f( \boldsymbol{Q}, t_0 )}{\Omega~}{\omega_k} \bigg)_{sc} = & ~ e^\frac{-\Delta E^2 t_{wid}^2}{16\ln{2}} 
\Bigg[ ~\bigg| a_\alpha \bra{\psi_f} e^{i\boldsymbol{Q} \cdot  \boldsymbol{r}} \ket{\psi_{\alpha}}  \bigg|^2  \\
&+ \bigg| a_\beta \bra{\psi_f} e^{i\boldsymbol{Q} \cdot  \boldsymbol{r}} \ket{\psi_{\beta}}  \bigg|^2 \\
 &+ 2 Re \bigg( a_\alpha^*
 \bra{\psi_f} e^{i\boldsymbol{Q} \cdot  \boldsymbol{r}} \ket{\psi_{\alpha}}^* \\
 & \times e^{i\phi(t_0)} a_\beta \bra{\psi_f} e^{i\boldsymbol{Q} \cdot  \boldsymbol{r}} \ket{\psi_\beta} \bigg) ~\Bigg].
\end{split}  
\end{equation}
The phase dependence in Fig.~\ref{Fig_Fix_mf} originates from the interference terms in Eq.(\ref{terms_wpkt_ddsp}). It is evident from the interference term in Eq.~(\ref{terms_wpkt_ddsp}) that when $a_\alpha$ and $a_\beta$ are real, $\phi = 0$ and $\phi = \pi$ cases depend on the real part of $\bra{\psi_f} e^{i\boldsymbol{Q} \cdot  \boldsymbol{r}} \ket{\psi_{\alpha}}^*\bra{\psi_f} e^{i\boldsymbol{Q} \cdot  \boldsymbol{r}} \ket{\psi_\beta}$ and the cases $\phi = \pi/2$ and $\phi = 3\pi/2$ depend on the imaginary part of this product. For the case in Fig.~\ref{Fig_Fix_mf}, from parity arguments $\bra{\psi_f} e^{i\boldsymbol{Q} \cdot  \boldsymbol{r}} \ket{\psi_{\alpha}}$ is real while $\bra{\psi_f} e^{i\boldsymbol{Q} \cdot  \boldsymbol{r}} \ket{\psi_{\beta}}$ is imaginary.  This leads to the product of the matrix elements being purely imaginary when $\phi = 0$ and $\phi = \pi$ thus making the $\phi = 0$ and $\phi = \pi$ plots in Fig.~\ref{Fig_Fix_mf} look identical. The overall shape of the plot (Fig.~\ref{Fig_Fix_mf}) however is largely determined by the non-interfering terms in Eq.(\ref{terms_wpkt_ddsp}).
One can expand the matrix elements that appear in these non-interfering terms, in a series of spherical harmonics $Y^{m'}_{l'}(\hat{\boldsymbol{Q}})$. For example for the parameters in Fig.~\ref{Fig_Fix_mf}, only the coefficients of  $Y^{0}_{2}(\hat{\boldsymbol{Q}})$ and $Y^{0}_{3}(\hat{\boldsymbol{Q}})$ are non-zero because of the rules associated with the addition of angular momentum. The coefficients of $Y^{0}_{2}(\hat{\boldsymbol{Q}})$ and $Y^{0}_{3}(\hat{\boldsymbol{Q}})$ involve an integral that depends on the radial part of the wave functions present and the spherical Bessel function $j_{2}(Qr)$ and $j_{3}(Qr)$ respectively which gives rise to the regions of minimum scattering probability seen in Fig.~\ref{Fig_Fix_mf}. These together give an idea of the shape of the plot in Fig.~\ref{Fig_Fix_mf}.

\subsection{Selecting the final state based on energy}\label{Sec_measurefinalmomentumonly}

Experimentally it might be reasonable to expect that the final states of the electron can only be broadly distinguished by their energies. Then, this would result in an incoherent sum of the scaled double differential scattering probability [Eq.~(\ref{wt_FT_equation})] over all the nearly-degenerate final states of the electron (sum over all possible $l$ and $m$ for a given $n$). The results shown in Fig.~\ref{Fig_sum_2s2p} involve a summation over the degenerate final states $2s$, 2$p_{-1}$, 2$p_{0}$, and 2$p_{1}$. Note that this is an incoherent sum over $\ket{\psi_f}$ as it involves the sum of the probabilities and not probability amplitudes. This is similar to the sum that appears in Ref.~\cite{Sim_2017}, except here the summation is only over the nearly-degenerate final states. It is evident from  Fig.~\ref{Fig_sum_2s2p}, that such a summation still preserves the dynamical phase information of the instantaneous wave packet. 

Since coincidental measurement of the final state of the electron and the momentum of the scattered photon have challenges, we discuss an alternative approach. If the scattered photons that correspond to the resonant transition to a specific final state of the electron can be precisely selected, then it would no longer be required to select the final state of the electron.
The idea of using energy-resolved measurements was discussed by Bennet et al.~\cite{Mukamel_Eresolved} to broadly distinguish between elastic and inelastic transitions. In this case, we discuss energy-resolved measurements as a way to precisely narrow down the scattered electron to a specific final state(s). As an extreme example, the average of the transition energy between 3$d\rightarrow$1$s$ and 4$f\rightarrow$1$s$ is $\sim$ 0.457 a.u. (12.44 eV) and the bandwidth of the incoming x-ray pulse (1 fs) is $\sim$ 0.055 a.u. (1.5 eV). If only those scattered photons with energy between 147.457 $\pm$ 0.028 a.u. are selected, then this effectively fixes the final state of the electron to be the 1$s$ state. The reason being that the transitions to other final states are unlikely given the limited bandwidth for the given scattered photon energy. This expectation is supported by calculations which show that the scaled double differential scattering probability for transitions to other final electron states are several orders of magnitude smaller than for the case of 1$s$. A detailed discussion of the energy-integrated double differential scattering probability is presented in the next section (Sec.~\ref{section_DDS_DS}). This approach is especially useful for selecting those final states which have a large energy separation from the wave packet constituent states and other neighbouring eigenstates. This technique can be exploited for several final states by choosing x-ray pulses with appropriate bandwidths. As an example, increasing the bandwidth of the x-ray pulse to 1.5 fs allows one to narrow the final state of the electron to be 2$s$ or one of the 2$p$ states without the need to detect it.

\subsection{Integrating the double differential scattering probability over the resolution of detector} \label{section_DDS_DS}
In this section, we discuss the effect of integrating the double differential scattering probability over the energy range of the detector. Typically, if the integration is performed over all possible energies, then it is referred to as the differential scattering probability. 

Consider Eq.~(\ref{diffprob_mainexpression}), for a desired electronic final state $\ket{\psi_f}$,  the detector is tuned to detect scattered photons centred around the frequency, 
\begin{equation}\label{Eqn_wkd}
  \omega_{kd} = \omega_{in} - E_f + E_{wpkt}.
\end{equation} 
Here $E_{wpkt}$ refers to the mean energy of the wave packet. The detector resolution is chosen to be $\delta$.

\begin{equation}\label{ds_step1}
  \pdv{P_f( \boldsymbol{Q}, t_0, \omega_{kd} )}{\Omega} = \frac{V}{(2\pi)^3} \int_{\omega_{kd} - \delta}^{\omega_{kd} + \delta} d\omega_k~ \omega^2_k ~ \pdv{P_f( \boldsymbol{Q}, t_0 )}{\Omega~}{\omega_k}.  
\end{equation}  
It should be noted that in the above expression we have assumed for simplicity that the detector window behaves like a step-function by only detecting the photons in the energy range $\omega_k \in [\omega_{kd} - \delta, \omega_{kd} + \delta]$.

First, the implications of the energy-integrated double differential scattering probability are discussed analytically with approximations. Then the integration is discussed numerically with an example. To proceed with the integration, the Waller-Hartree approximation~\cite{waller_hartree_1929} is applied to Eq.~(\ref{ds_step1}) which refers to the assumption, $\omega_k \approx \omega_{in}$ and $\boldsymbol{Q}$ is largely independent of $\omega_k$, Eq.~(\ref{ds_step1}). This approximation is valid as long as the energy transferred by the x-ray probe photon to the scattered electron is small compared to $\omega_{in}$ ~\cite{Sim_2017}. This leads to,

\begin{equation}\label{ds_step2}
\begin{split}
    \pdv{P_f( \boldsymbol{Q}, t_0, \omega_{kd} )}{\Omega} =& \frac{V~\omega^2_{in}}{(2\pi)^3}  \pdv{P_e}{\Omega~}{\omega_k} \\
    & \times \sum\limits_{ n', n'' } a^*_{n'} a_{n''} ~e^{i(E_{n'} - E_{n''} ) t_0} \\
    & \times \bra{\psi_{n'}} e^{-i \boldsymbol{Q} \cdot \boldsymbol{r} }  \ket{\psi_{f}}  \bra{\psi_{f}} e^{i \boldsymbol{Q} \cdot \boldsymbol{r} }  \ket{\psi_{n''}} \\
    & \times \int_{\omega_{kd} - \delta}^{\omega_{kd} + \delta} d\omega_k e^{-\big[(\epsilon_f - E_{n'})^2 + (\epsilon_f - E_{n''})^2 \big]  \frac{t^2_{wid}}{8\ln{2}}  }
\end{split}
\end{equation} 
Examining Eq.~(\ref{ds_step2}) for the case of a two-state wave packet, there are two integrals from the non-interference terms and an integral that arises from the interference term.

\begin{equation}\label{ds_step3}
\begin{split}
\bigg( \pdv{P_f( \boldsymbol{Q}, t_0, \omega_{kd} )}{\Omega} \bigg)_{sc} = & ~
\Bigg[ I_1\bigg| a_\alpha \bra{\psi_f} e^{i\boldsymbol{Q} \cdot  \boldsymbol{r}} \ket{\psi_{\alpha}}  \bigg|^2  \\
&+ I_2\bigg| a_\beta \bra{\psi_f} e^{i\boldsymbol{Q} \cdot  \boldsymbol{r}} \ket{\psi_{\beta}}  \bigg|^2 \\
 &+ 2 I_3~Re \bigg( a_\alpha^*
 \bra{\psi_f} e^{i\boldsymbol{Q} \cdot  \boldsymbol{r}} \ket{\psi_{\alpha}}^* \\
 & \times e^{i\phi(t_0)} a_\beta \bra{\psi_f} e^{i\boldsymbol{Q} \cdot  \boldsymbol{r}} \ket{\psi_\beta} \bigg) ~\Bigg].
\end{split}  
\end{equation}
where $I_1$ and $I_2$ are the integrals that arise for the non-interference terms and $I_3$ is the integral for the interference term. In the left hand side of the above equation [Eq.~(\ref{ds_step3})], the energy-integrated double differential scattering probability has been scaled by the free electron expression similar to the earlier convention for double differential scattering probability~[Eq.~(\ref{terms_wpkt_ddsp})].

\begin{equation}\label{integral1_ddsp}
\begin{split}
I_1 &= \int_{\omega_{kd} - \delta}^{\omega_{kd} + \delta} d\omega_k e^{-\big[(E_f + \omega_k - \omega_{in} - E_{\alpha} )^2  \big]   2\tau  } \\
   &= \frac{1}{2} \sqrt{\frac{\pi}{2\tau}} \Bigg[ erf\bigg( \sqrt{2\tau} \bigg[ \frac{1}{2}(E_\alpha - E_\beta) + \delta \bigg] \bigg) \\
   & ~ ~ ~ - erf\bigg(  \sqrt{2\tau} \bigg[ \frac{1}{2}(E_\alpha - E_\beta) - \delta \bigg]  \bigg)  \Bigg].
\end{split}  
\end{equation}

\begin{equation}\label{integral2_ddsp}
\begin{split}
I_2 &= \int_{\omega_{kd} - \delta}^{\omega_{kd} + \delta} d\omega_k e^{-\big[(E_f + \omega_k - \omega_{in} - E_{\beta} )^2  \big]  2\tau  } \\
&= \frac{1}{2} \sqrt{\frac{\pi}{2\tau}} \Bigg[ erf\bigg( \sqrt{2\tau} \bigg[ \frac{1}{2}(E_\beta - E_\alpha) + \delta \bigg] \bigg) \\
  & ~ ~ ~ - erf\bigg(  \sqrt{2\tau} \bigg[ \frac{1}{2}(E_\beta - E_\alpha) - \delta \bigg]  \bigg)  \Bigg].
\end{split}  
\end{equation}

\begin{equation}\label{integral3_ddsp}
\begin{split}
I_3 &= \int_{\omega_{kd} - \delta}^{\omega_{kd} + \delta}  d\omega_k \exp \Bigg( -\tau \bigg[ (E_f + \omega_k - \omega_{in} - E_{\alpha} )^2 \\ 
 & ~ ~ ~ ~ +(E_f + \omega_k - \omega_{in} - E_{\beta} )^2  \bigg] \Bigg)   \\
 &= \frac{1}{2} \sqrt{\frac{\pi}{2\tau}} e^{\frac{-\tau}{2} (E_\beta - E_\alpha)^2 } \Bigg[ erf\big( \sqrt{2\tau} \delta \big) \\ 
 & ~ ~ ~ - erf\big(-\sqrt{2\tau} \delta   \big)  \Bigg].
\end{split}  
\end{equation}
In the above equations, we have used $\omega_{kd} =  \omega_{in} - \big(E_f - \frac{1}{2} ( E_{\alpha} + E_{\beta} ) \big)$. The quantity $\tau = \frac{t^2_{wid}}{8\ln{2}}$.

\begin{figure}
\resizebox{80mm}{!}{\includegraphics{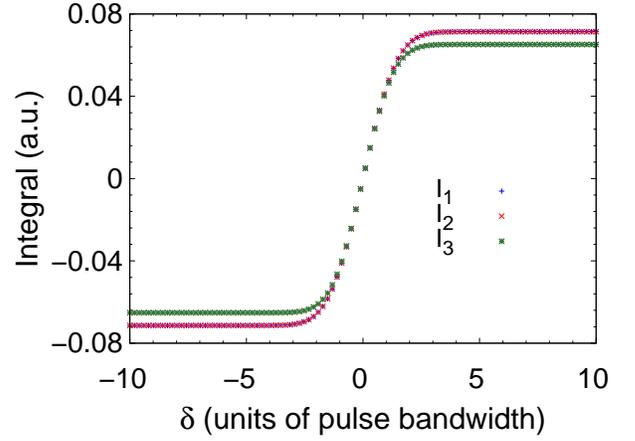}}
\caption{\label{Fig_plotintegrals}
The plots show the integrals in Eqs.~(\ref{integral1_ddsp}) - (\ref{integral3_ddsp}) as a function of detector resolution $\delta$ which is expressed in multiples of x-ray probe pulse energy bandwidth. It is evident that the integrals that come from integrating the non-interference terms $I_1$ amd $I_2$ are equal. Also, the integral $I_3$ that arises from integrating the interference term of the double differential scattering probability is nearly equal to $I_1$ even for large detector resolutions ($\delta$). Therefore, the energy-integrated double differential scattering probability is still nearly proportional to the double differential scattering probability. The parameters used are the same as that of Fig.~\ref{Fig_Fix_mf} except here the final state is chosen to $1s$.
}
\end{figure}

A comparison of the integrals (see Fig.~\ref{Fig_plotintegrals}) from Eqs.~(\ref{integral1_ddsp}), (\ref{integral2_ddsp}), and (\ref{integral3_ddsp}) shows that $I_1$ = $I_2$. The integrals $I_1$ and $I_3$ agree for small detector resolutions ($\delta$) and are nearly equal (to within a few percent) for detector resolutions ($\delta$) that are as high as ten times the bandwidth of the pulse. This implies that the energy -integrated double differential scattering probability~[Eq.~(\ref{ds_step3})] is still nearly proportional to the double differential scattering probability~[Eq.~(\ref{terms_wpkt_ddsp})]. It should be noted that the effect of integrating the double differential scattering probability can be precisely broken down. That is the non-interference terms reveal the effects of $I_1$ and $I_2$ and the interference terms reveal the effects of $I_3$. Remember that the interference term depends on the probe-delay time [Eq.~(\ref{terms_wpkt_ddsp})] while the non-interference terms do not. Therefore, the effect of these integrals can be extracted experimentally in principle by examining the interference and the non-interference terms of the energy-integrated double differential scattering probability. 


Independent of the analytic discussion, we now calculate the energy-integrated double differential scattering probability [Eq.~(\ref{ds_step1})] numerically for the case discussed in Sec.~\ref{Sec_measurefinalmomentumonly} that is when $\ket{\psi_f} = 1s$. For this case, $\omega_{kd} = 147.457$ a.u. Let $\delta = 0.22$ a.u. (6 eV). This corresponds to a detector resolution that is eight times the bandwidth of the pulse. To calculate the energy-integrated double differential scattering probability numerically, it is assumed that $\omega_k \approx \omega_{in}$ in the $\omega^2_k$ term on the right-hand side of Eq.~(\ref{ds_step1}). However unlike in the analytic discussion, for the integration, $\boldsymbol{Q}$ in $\pdv{P_f( \boldsymbol{Q}, t_0 )}{\Omega~}{\omega_k}$ is not assumed to be independent of $\omega_k$.  Despite this, we find that

\begin{equation}\label{ds_numericalstep1}
  \pdv{P_f( \boldsymbol{Q}, t_0, \omega_{kd}, \delta )}{\Omega} \propto \pdv{P_f( \boldsymbol{Q}, t_0 )}{\Omega~}{\omega_k}.
\end{equation} 
Here the double differential scattering probability was evaluated at $\omega_k = \omega_{kd}$. while varying $\boldsymbol{Q}$ and $t_0$. The deviation from the above proportionality ($\sim 2\%$) is lower than what is expected from Fig.~\ref{Fig_plotintegrals}. The reason is the double differential scattering probability is suppressed by orders of magnitude when $\omega_{k}$ is far from $\omega_{kd}$, so these values don't contribute to the energy-integrated expression as much. Given the proportionality, the double differential scattering probability can be used directly instead of the energy-integrated double differential scattering probability to understand the dynamics of the wave packet.

It should be noted that in the above example, the chosen resolution of the detector~(eight times the bandwidth of the pulse) is much less stringent than the typical resolutions used for the same problem previously~\cite{Dixit_mainpnas, Sim_2017}. For instance in Ref.\cite{Sim_2017} a detector energy resolution of 0.25 eV which is 1/3 of the pulse energy bandwidth is used. The reason we don't require such high resolutions is because in the examples we discussed, the scattering has more inelastic character than the transitions examined in Ref.~\cite{Dixit_mainpnas, Sim_2017}. This is an important point, because it is this inelastic behaviour which allows us to select the final state(s). Hence making it possible to interpret the dynamics from the scattering signal which was found to be difficult in Ref.~\cite{Dixit_mainpnas}. However, there is a trade-off in that inelastic transitions have a lower overall probability than elastic ones, thus leading to a lower signal strength than in the previous cases. For comparison, the peak of the signal in the case of transition to 1s ($\omega_k = 147.457 \pm 0.22 $ a.u.) is more than two orders of magnitude smaller than the peak of the elastic signal ($\omega_k = 147 \pm 0.0092 $ a.u.). However, if we consider the same detector resolution as that of Ref.~\cite{Dixit_mainpnas, Sim_2017}, there is a more interesting case. The inelastic transition where the final state is either 2s or 2p ($\omega_k = 147.082 \pm 0.0092  $ a.u.) gives a peak signal which is only an order of magnitude smaller than the elastic signal ($\omega_k = 147 \pm 0.0092 $ a.u.). Here, the inelastic transition to 2s or 2p constitutes roughly 90\% of the signal. Therefore, in this case the scattering pattern can be interpreted to reveal the instantaneous transition charge density which would have been difficult in the elastic case examined in Refs.~\cite{Dixit_mainpnas, Sim_2017} and this is achieved by a mere detuning of the detector.

If one needs to access the instantaneous transition charge density from the x-ray scattering profile using Eq.~(\ref{wt_FT_equation}), either one has to have a detector resolution in the range discussed above or one can resort to coincidental measurement wherein the final state of the scattered electron is fixed and the scattered photon momentum is measured simultaneously. However, the use of coincidental measurements offers much more flexibility with detector resolution. For instance when the final state of the electron is 1s, most of the transition probability is captured by the scattered photons that are within a couple of bandwidths from $\omega_{kd}$. 
If the scattered photons that are far away in energy from $\omega_{kd}$ are detected, in an ideal coincidental measurement only photons that are coming from the electronic transition to final state 1s are counted. Effectively, in the case of coincidental measurements, one has a chance to trade-off accuracy in the detector resolution of the scattered photon energy, with accuracy in the coincidental measurement of the final state of the electron. As a limiting case, one can think of the converse scenario where a highly accurate measurement of the final state of the electron can be substituted for any measurement of the scattered photon energy but note that the direction and the scattered photon count are still required.

So far the discussions have focused around a two-state wave packet but the main results presented in this work are valid for an arbitrary wave packet. In this spirit, we perform calculations for the case of a wave packet consisting of a superposition of three eigenstates. The wave packet was chosen to consist of equal probabilities of $3d$, $4f$ and $5p$. When the final state of the electron is fixed to be 1s, unsurprisingly it was found that Eq.~(\ref{ds_numericalstep1}) still holds true for similar detector resolutions.

\subsection{Case of a partially known wave packet}
As a final example, the case of a wave packet [Eq.~(\ref{wpkt_defn})] which is largely made up of a known eigenstate $\ket{\psi_{\beta}}$ with $a_\beta \simeq 1$ and a small amount $a_\alpha$ of an unknown eigenstate $\ket{\psi_\alpha}$ is explored. We offer a method using the approach described in Sec.~\ref{Methods} to determine the unknown eigenstate. 


\begin{figure}
\resizebox{65mm}{!}{\includegraphics{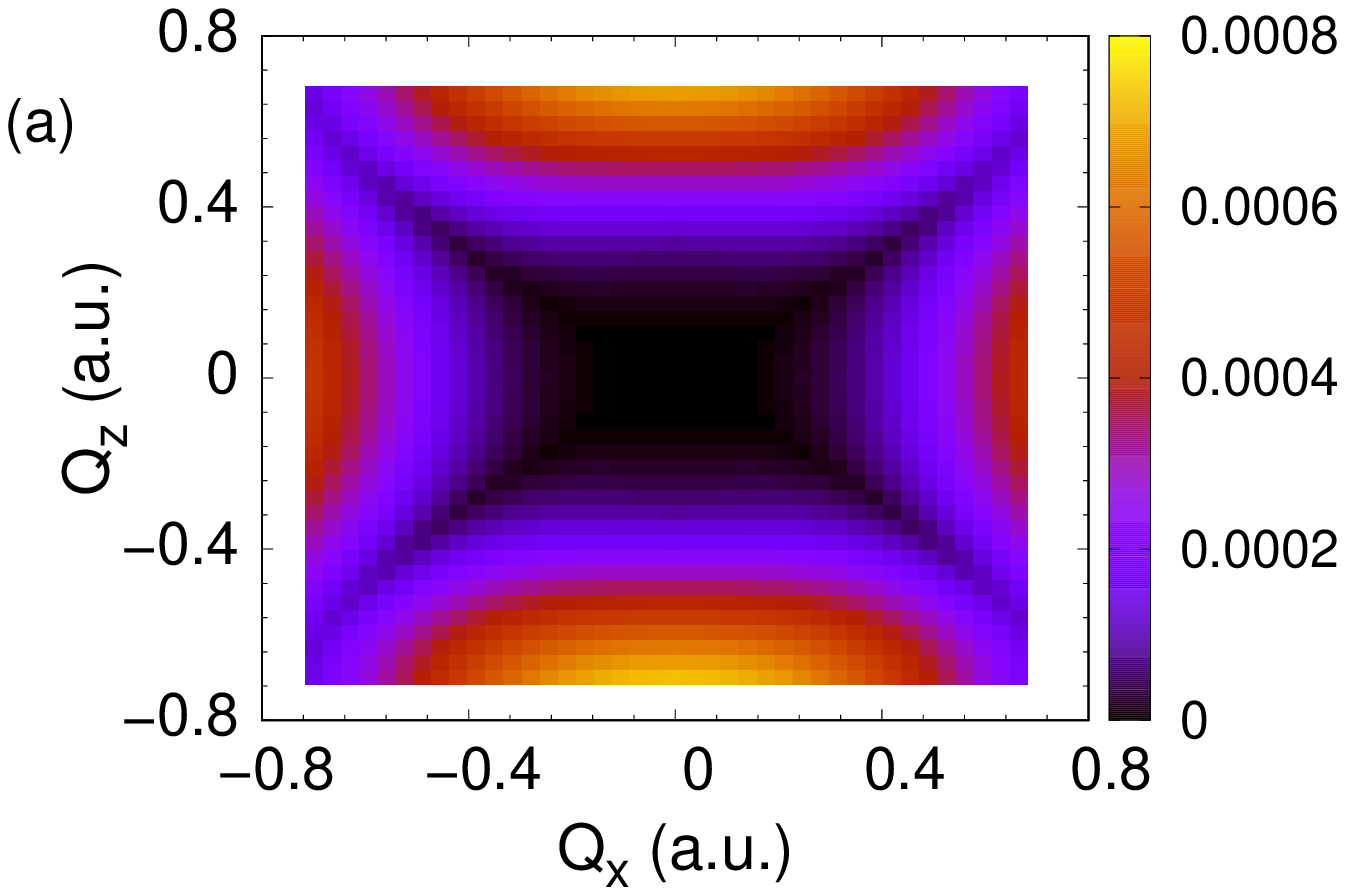}}
\resizebox{65mm}{!}{\includegraphics{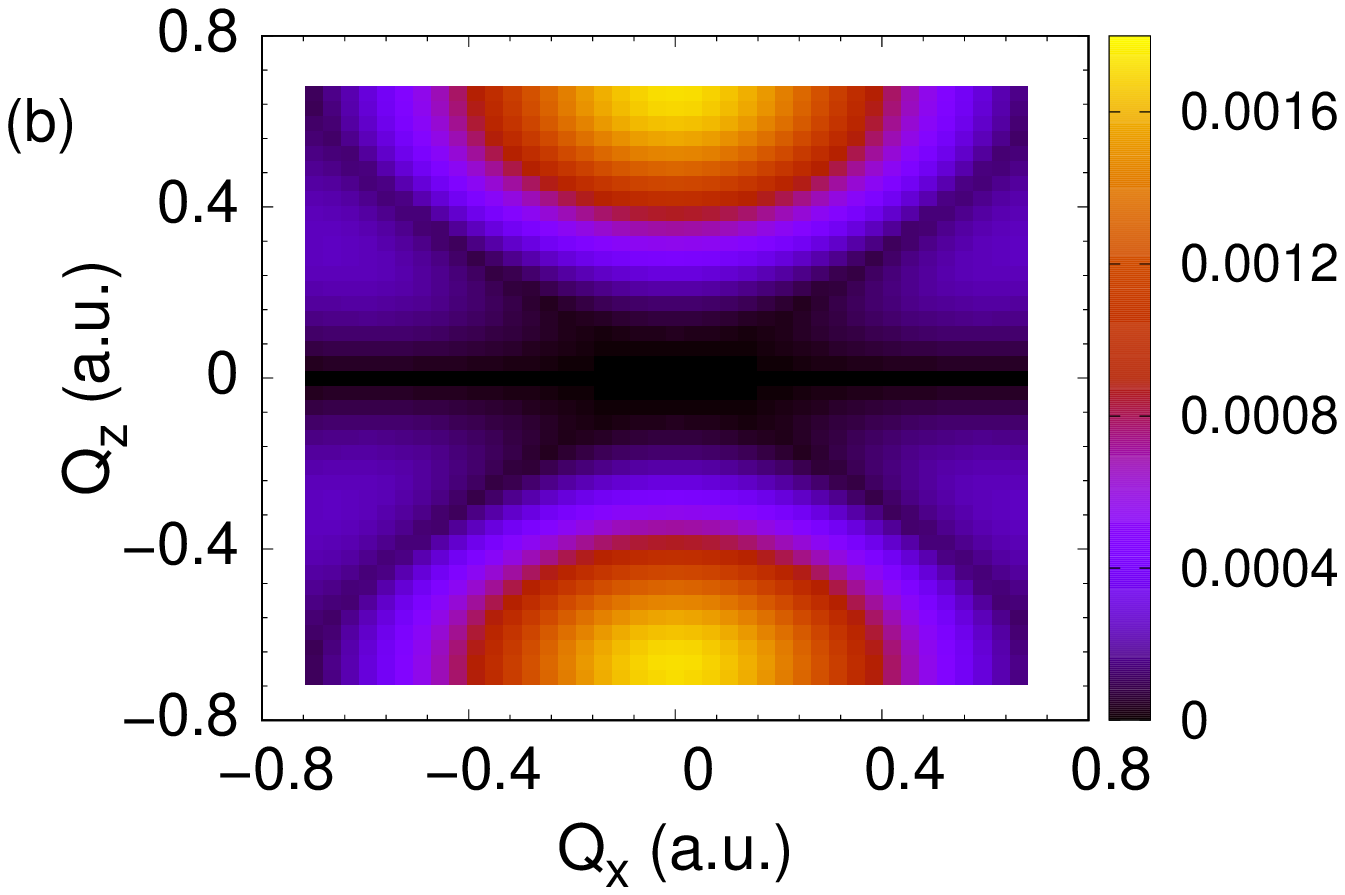}}
\resizebox{65mm}{!}{\includegraphics{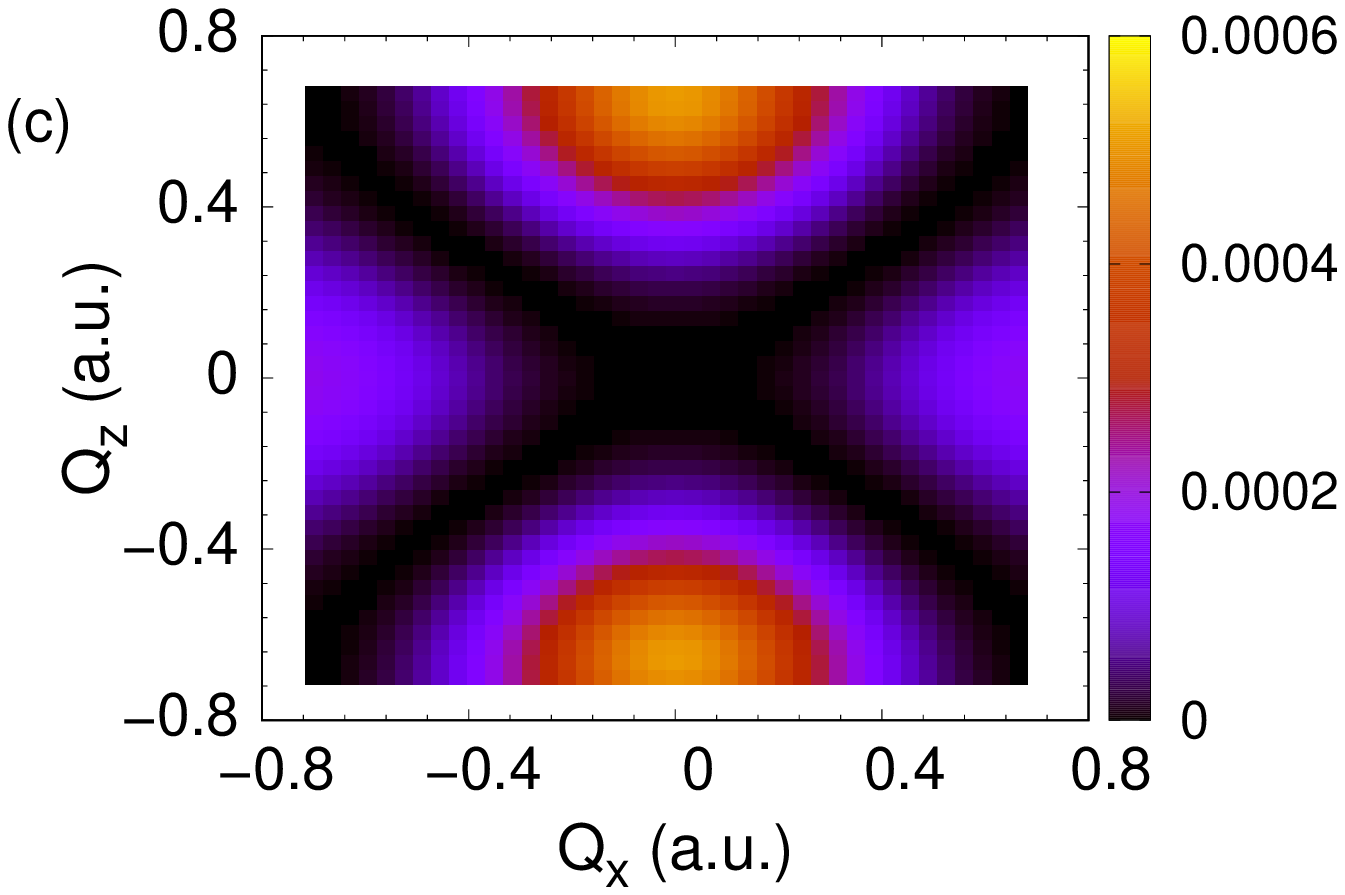}}
\resizebox{65mm}{!}{\includegraphics{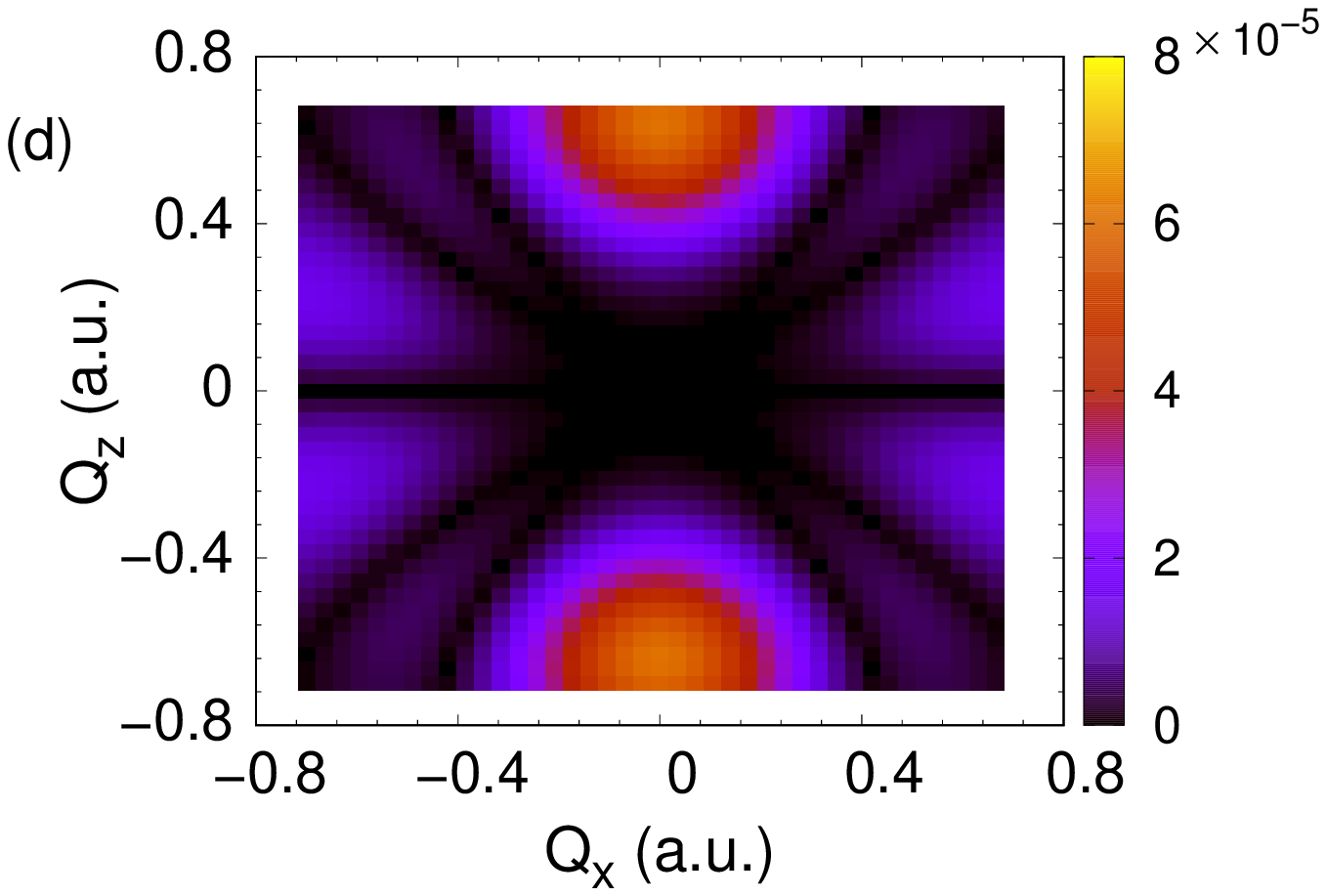}}
\caption{\label{Fig_unknownwpkt_varyl} 
The results for the amplitude of the interference term [Eq.~(\ref{amplitude_expression})] vs momentum transferred to the electron. Here the final state of the electron is chosen to be $1s$. The initial wave packet consists of 95\% of $\ket{\psi_\beta} = \ket{3d_0}$ and 5\% probability of an unknown eigenstate $\ket{\psi_\alpha}$. Different cases for the unknown state $\ket{\psi_\alpha}$ are explored with (a) $4s$ (b) $4p_0$, (c)$4d_0$, and (d) $4f_0$. It is evident that the choice of the initial wave packet leaves a finger print on the x-ray scattering profile. This can be used to uniquely identify the unknown eigenstate in the initial wave packet. The other parameters are the same as Fig.~\ref{Fig_Fix_mf}. A qualitative way to understand the decreasing spread in momentum space from plots (a)-(d) is from the uncertainty principle. The amplitude plotted involves matrix elements using the state $\ket{\psi_\alpha}$  whose uncertainty in position increases from plots (a)-(d) as the orbital angular momentum $l$ increases for a given principal quantum number $n$.
}
\end{figure}

\begin{figure*} 
\resizebox{175mm}{!}{\includegraphics{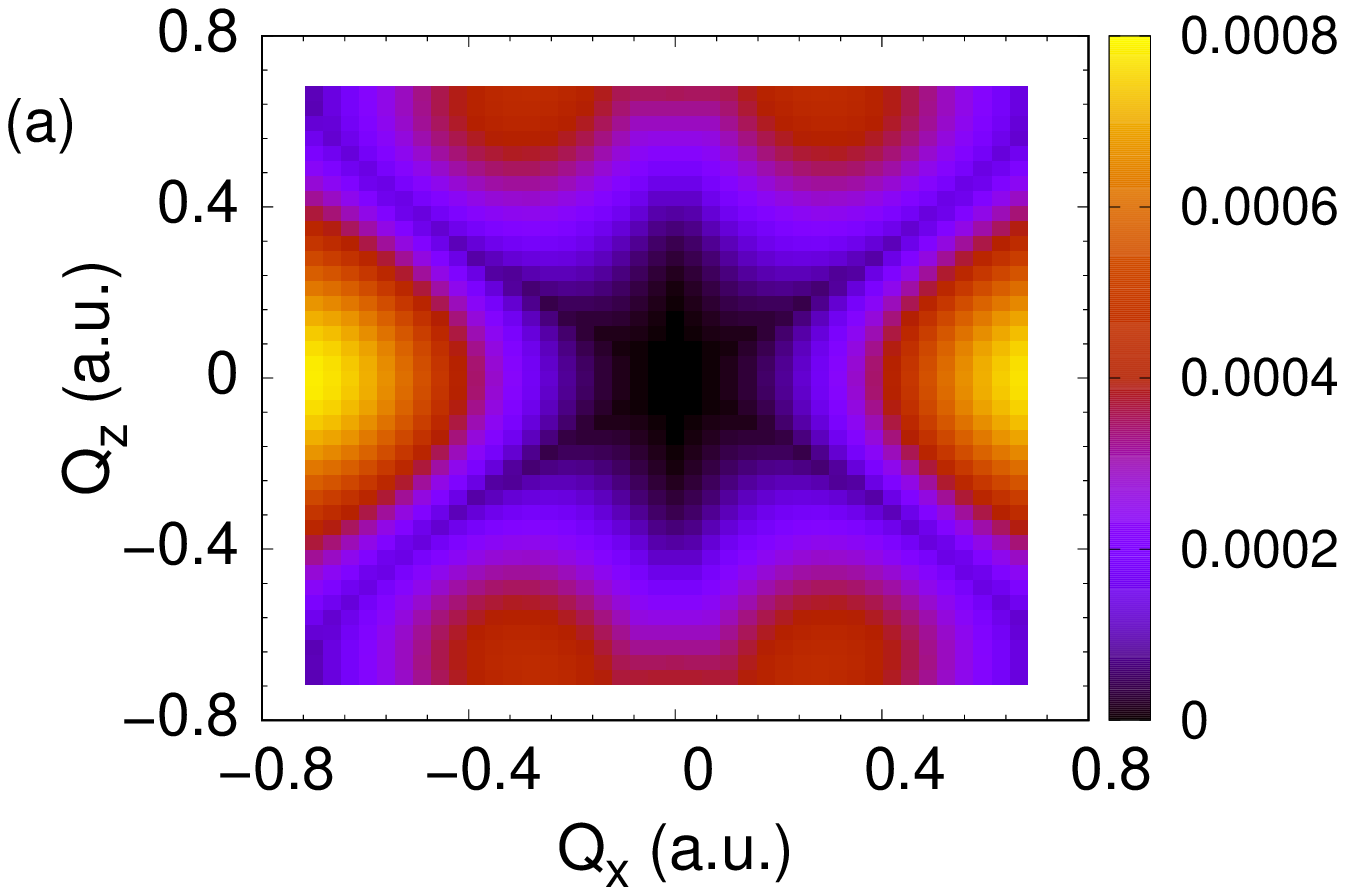} \includegraphics{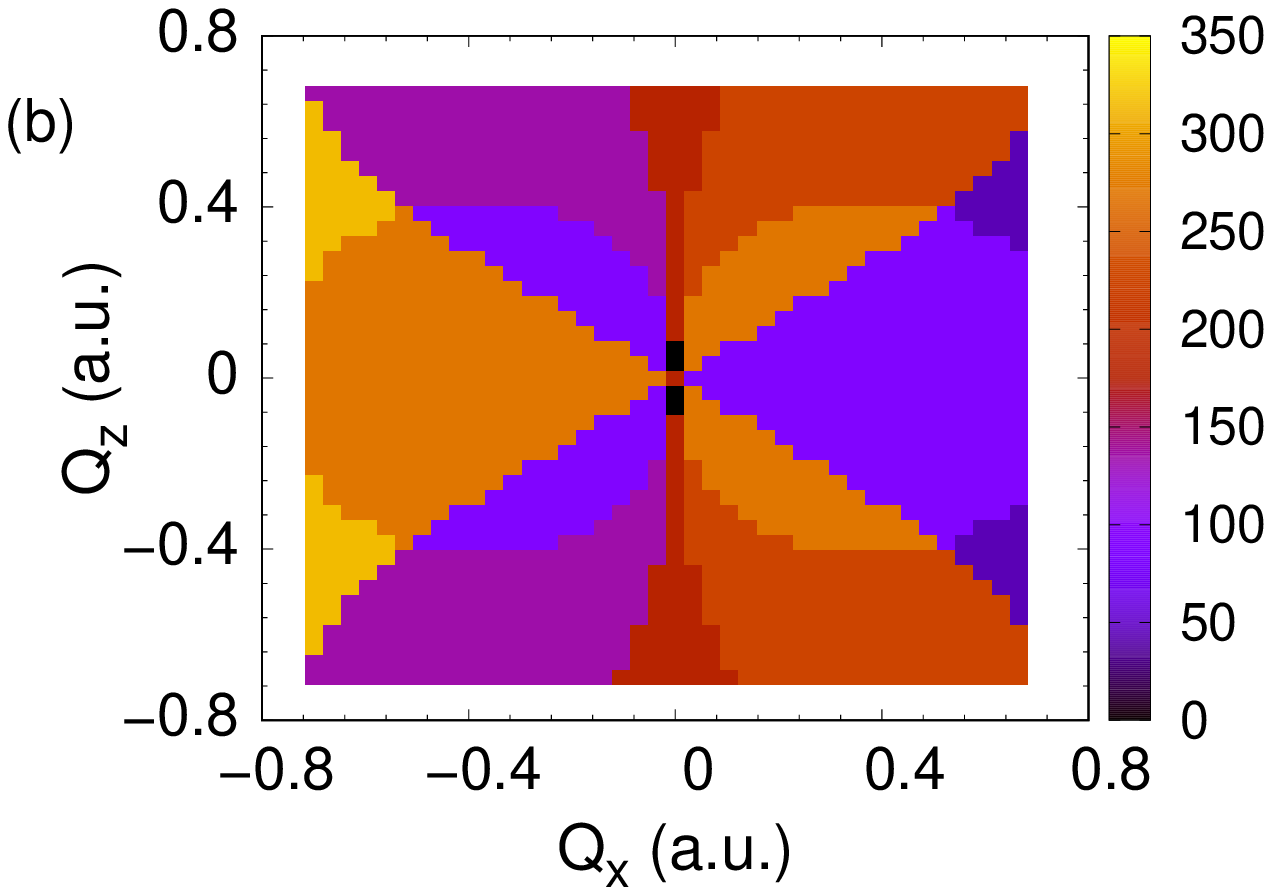} \includegraphics{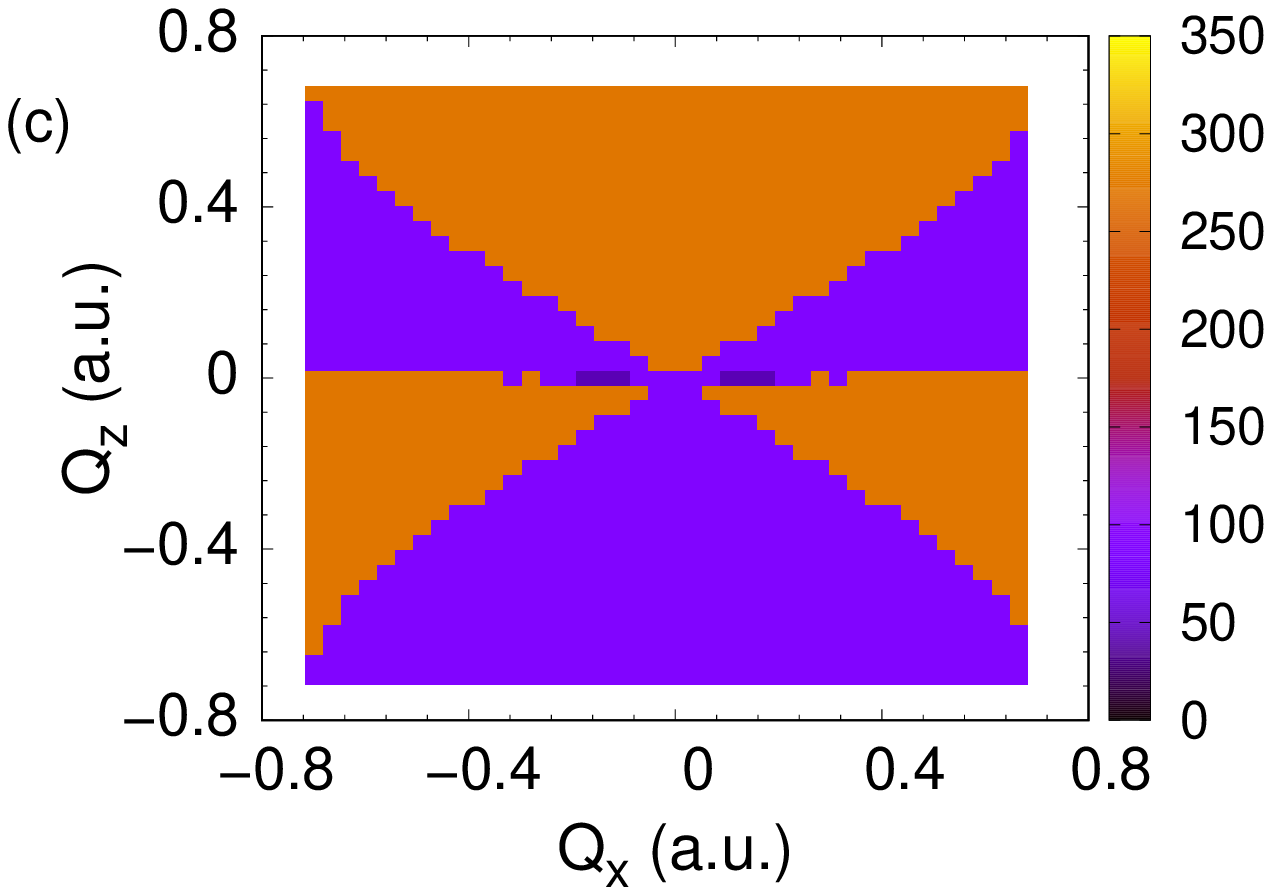} \includegraphics{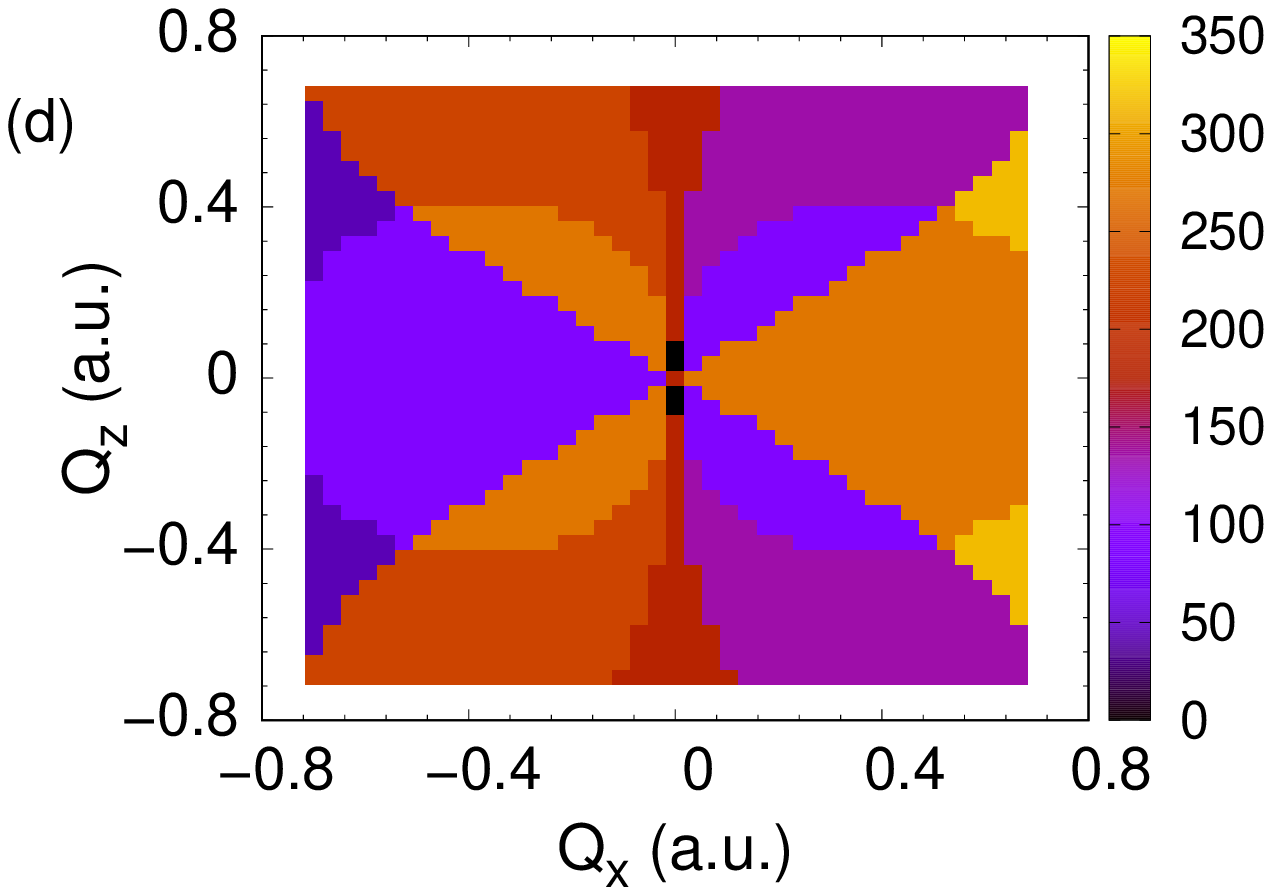} }
\caption{\label{Fig_unknownwpkt_varym} 
The effect of the magnetic quantum number $m$ of the eigenstates in the wave packet on the amplitude and the intrinsic phase of the interference term. Plot (a) contains the amplitude of the interference term [Eq.~(\ref{amplitude_expression})] vs momentum transferred to the electron. The initial wave packet consist of 3d and $4p_1$ states with all the other parameters being the same as Fig.~\ref{Fig_unknownwpkt_varyl}. Note that the amplitude of the interference term is only sensitive to the absolute value of the magnetic quantum number $|m|$ of the eigenstates in the initial wave packet (compare with Fig.~\ref{Fig_unknownwpkt_varyl}(b) ). Plots (b) - (d) reveal the dependence of the intrinsic phase $\delta$ (in degrees) of the interference term on the momentum transferred to the electron. The unknown eigenstate $\ket{\psi_\alpha}$ is chosen to be (b) $4p_{-1}$, (c) $4p_{0}$, and (d) $4p_{1}$.
}
\end{figure*}

In the previous examples, the time-dependence of the scattering profile was presented and could be seen to originate from the time-dependent interference terms. In this case, we present the time-independent amplitude of the interference terms which reveals  properties of the unknown eigenstate. This quantity can be calculated by examining the terms in Eq.~(\ref{terms_wpkt_ddsp}). The first term can be neglected given that $a_\alpha$ is small. The second term in Eq.~(\ref{terms_wpkt_ddsp}) is a known quantity. The interference (third) term before evaluating the real part can be written as $\chi e^{i(\phi(t_0) - \delta)}$ for some real $\chi$ and intrinsic phase $\delta$ which depend on the matrix elements. For a given $\boldsymbol{Q}$, the amplitude of the interference term ($\chi$) can be calculated from Eq.~(\ref{terms_wpkt_ddsp}) by varying $\phi(t)$ to obtain the maximum value. Algebraically one can show
\begin{equation}\label{amplitude_expression}
\begin{split}
  \chi (\boldsymbol{Q}) = & ~ 2e^\frac{-\Delta E^2 t_{wid}^2}{16\ln{2}} ~\Bigg|  a_\alpha \bra{\psi_f} e^{i\boldsymbol{Q} \cdot  \boldsymbol{r}} \ket{\psi_{\alpha}} \\
  &\times a_\beta \bra{\psi_f} e^{i\boldsymbol{Q} \cdot  \boldsymbol{r}} \ket{\psi_\beta} \Bigg|.
 \end{split} 
\end{equation}
The amplitude of the interference term along with the intrinsic phase ($\delta$) serve as a fingerprint of the eigenstates in the initial wave packet (Figs.~\ref{Fig_unknownwpkt_varyl} and \ref{Fig_unknownwpkt_varym}). The amplitude plots alone shown in Fig.~\ref{Fig_unknownwpkt_varyl} and Fig.~\ref{Fig_unknownwpkt_varym}(a) reveal substantial qualitative differences which can be used to uniquely identify the unknown state up to a given $|m|$ value. To distinguish between the different signs of the magnetic quantum number $m$ for the unknown state, one can examine the plots in Fig.~\ref{Fig_unknownwpkt_varym} describing the dependence of the intrinsic phase ($\delta$) on $\boldsymbol{Q}$.

In principle, experimentally one can estimate the unknown eigenstate using the following steps. First, the energy of the unknown eigenstate can be determined from the time period of oscillation of the wave packet. This can be measured from the x-ray scattering profile by varying the delay time $t_0$ with no measurement of the final state of the electron required. Recall that the time period of oscillation of the wave packet is $2\pi/|E_\alpha - E_\beta|$.  Given the energy of the unknown eigenstate in the wave packet, one can extract the amplitude ($\chi$) of the interference term by making successive measurements of the scaled double differential scattering probability at different delay times $t_0$ for the wave packet $\psi(\boldsymbol{r},t_0)$. For a given $\boldsymbol{Q}$, the delay time that results in the largest magnitude of the interference term can be used to obtain the intrinsic phase. The amplitude ($\chi$) of the interference term and the intrinsic phase ($\delta$) profile (see Figs.~\ref{Fig_unknownwpkt_varyl} and \ref{Fig_unknownwpkt_varym}) can then be used to identify the unknown state from a set of eigenstates of the system.

For the examples discussed in Figs.~\ref{Fig_unknownwpkt_varyl} and \ref{Fig_unknownwpkt_varym} the choice for the final state of the electron to be $1s$ may appear to be challenging because of it being the ground state. However, if one follows the approach discussed in Sec.~\ref{Sec_measurefinalmomentumonly} one needs to measure only the scattered photon momentum precisely to obtain the scattering profiles without any need for coincidental measurements of the final state of the electron .


\section{Conclusion and summary} \label{conclusion_summary}
Previous research on x-ray scattering from a wave packet revealed that the scattering patterns have a non-trivial dependence on the instantaneous probability density of the wave packet~\cite{Dixit_mainpnas}. In this work, we discussed how coincidentally selecting the final state of the scattered electron and the momentum of the scattered photon allows one to extract information about the instantaneous probability density of the wave packet from the scattering signal. The double differential scattering probability from the wave packet was found to be proportional to the modulus square of the Fourier transform of the instantaneous transition charge density. An alternative method which only requires a measurement of the scattered photon momentum without the need to simultaneously measure the final state of the electron was also presented. It was shown to be applicable in cases where the scattered photon energy can be measured precisely enough such that only the transition from the wave packet states to the desired final state(s) occurs. The effect of the energy resolution of the photon detector on the scattering probability was presented. Several examples were discussed with an emphasis on cases that might be more experimentally favourable. Finally, the case of x-ray scattering from a wave packet which is largely (e.g.~$95\%$ probability) made of a known eigenstate and has a small amount ($5\%$) of an unknown eigenstate is discussed. The amplitude of the interference term in the double differential scattering probability and its intrinsic phase can be used to identify the unknown eigenstate. 

It is worth pointing out that for all of the examples explored in this work, strong incident fields while not necessary (Intensity~$\sim$~$10^{20}$ W/$\text{cm}^2$) can be used to obtain a larger absolute differential scattering probability if desired. From a theoretical perspective,  Eq.~(\ref{diffprob_mainexpression}) is valid even for x-ray intensities $\sim$~$10^{20}$ W/$\text{cm}^2$. A more detailed discussion on the validity of the perturbative approach in the strong field regime can be found in Ref.~\cite{NLC_interference}. From an experimental perspective, it should be noted that the currently available XFELs~\cite{Fuchs,LCLS_specswebsite, SACLA_specs_website1, SACLA_specs_website2, EuropeanXFEL_chem_specs,EuropeanXFEL1,EuropeanXFEL2,EuropeanXFEL3} are capable of generating x rays in the discussed parameter regime.

\section{Acknowledgements}
This work was supported by the U.S. Department of Energy, Office of
Science, Basic Energy Sciences, under Award No. DE-SC0012193. We are extremely grateful to N.H. Shivaram for discussions on x-ray scattering and the experimental challenges involved.

\bibliography{References.bib}

\end{document}